\documentclass[twocolumn]{aa}
\usepackage[usenames]{color}
\usepackage{mathrsfs}
\usepackage{graphicx}
\usepackage{natbib}
\usepackage{enumitem}
\usepackage{hyperref}
\usepackage[normalem]{ulem}
\usepackage{amsmath}
\usepackage{graphicx}

\newcommand{\kms}{{\mathrm{\, km \,s^{-1}}}}

\newcommand\width{0.4}

\newcommand{\clustersize}{N_\mathrm{adjacent}}
\newcommand{\avesigma}{\left<\sigma\right>}
\newcommand{\arcsecs}{\, \mathrm{arcsecs}}

\DeclareGraphicsExtensions{.pdf,.png,.jpg}

\newcommand{\mins}{{\mathrm{\, minutes}}}

\definecolor{manuel}{rgb}{0.80, 0.4, 0.4}
\definecolor{joan}{rgb}{0.93, 0.57, 0.13}
\definecolor{frederic}{rgb}{0.0,0.5,0.0}
\definecolor{valeriia}{rgb}{0.0,0.5,0.0}
\definecolor{terry}{rgb}{0.0,0.5,0.0}

\begin{document}
\title{Automatic detection technique for solar filament oscillations in GONG data.}
\author{authors}


\date{Received <date> /
Accepted <date>}

\author{M. Luna\inst{1,2}
	J. R. M\'erou Mestre\inst{1,2}
	\and
	F. Auchère\inst{3}
     }

  \institute{Departament F{\'i}sica, Universitat de les Illes Balears, E-07122 Palma de Mallorca, Spain
       \email{manuel.luna@uib.es}
       \and
       Institute of Applied Computing \& Community Code (IAC$^3$), UIB, Spain
       \and
       Université Paris-Saclay, CNRS, Institut d’Astrophysique Spatiale, 91405 Orsay, France  }

\titlerunning{Detection}
\authorrunning{Luna, M\'erou Mestre \& Auchère}

\abstract{Solar filament oscillations have been known for decades. Now thanks to the new capabilities of the new telescopes, these periodic motions are routinely observed. Oscillations in filaments show key aspects of their structure. A systematic study of filament oscillations over the solar cycle can shed light on the evolution of the prominences.}
{This work is a proof of concept that aims to automatically detect and parameterise such oscillations using H$\alpha$ data from the GONG network of telescopes.}
{The proposed technique studies the periodic fluctuations of every pixel of the H$\alpha$ data cubes. Using the FFT we compute the power spectral density (PSD). We define a criterion to consider whether it is a real oscillation or whether it is a spurious fluctuation. This consists in considering that the peak in the PSD must be greater than several times the background noise with a confidence level of 95\%. The background noise is well fitted to a combination of red and white noise. We applied the method to several observations already reported in the literature to determine its reliability. We also applied the method to a test case, which is a data set in which the oscillations of the filaments were not known a priori.}
{The method shows that there are areas in the filaments with PSD above the threshold value. The periodicities obtained are in general agreement with the values obtained by other methods. In the test case, the method detects oscillations in several filaments.}
{We conclude that the proposed spectral technique is a powerful tool to automatically detect oscillations in prominences using H$\alpha$ data.}


\maketitle

\section{Introduction}\label{sec:intro}

Solar filaments (also called prominences when seen off-disk) are clouds of dense and cold plasma levitating in the tenuous and hot corona supported against gravity by their magnetic field. In H$\alpha$, they are seen in absorption as dark structures on the solar disk. 
There are still many unknowns about their nature: how they form, their structure or how they disappear. Prominences form in large magnetic structures called filament channels (FC).
These structures are highly non-potential fields appearing above the polarity inversion lines (PIL) of the photosphere. The observational evidence suggests that the FCs are produced by the interaction of the solar magnetic field with the surface and subsurface motions of the sun. These motions are a combination of the large-scale differential rotation, meridional flows, surface diffusion and smaller-scale vortical motions such as granular and supergranular convection \citep[see review by][]{mackay_formation_2014}. 
During the solar cycle, filaments appear in different latitudes forming a pattern. They migrate over a range of latitudes \citep{mcintosh_solar_1972,minarovjech_prominences_1998} showing a butterfly diagram \citep{coffey_digital_1998} having similarities to the sunspot diagram. This pattern indicates that the structure of the filaments may be related to the large-scale solar field and evolve over the solar cycle.

Solar prominences are structures subject to a wide variety of movements, including oscillations. These periodic motions have been observed sporadically since the beginning of the 20th century, but thanks to new instruments, observations of filament oscillations have become common \cite[see review by][]{arregui_prominence_2018}. 
The relevance of these periodic motions is that we can apply prominence seismology. 
This remote diagnostics method aims to determine physical parameters that are difficult to measure by direct means in these magnetized plasma structures. 
It combines observations of oscillations with theoretical results from the analysis of oscillatory properties of given prominence models. 
It is necessary to discriminate the restoring force causing the oscillations to apply seismology. 
The relative direction of the plasma motion to the magnetic field determines which combination of forces is involved. The oscillations can be grouped into transverse where the restoring force is mostly the Lorentz force \citep[e.g.][]{hyder_winking_1966,kleczekkuperus_oscillatory_1969,hershaw_damped_2011} and longitudinal where the restoring force is mostly gravity along the field lines together with the gas pressure gradients \citep[e.g.][]{zhang_observations_2012,luna_observations_2014,bi_solar_2014}.
\citet{luna_large-amplitude_2017} compared those field properties derived from filament seismology with the characteristics of an inserted flux rope \citep{van-ballegooijen_observations_2004} and found them to be in good general agreement.

Oscillations in filaments show key aspects of their structure. A systematic study of filament oscillations over the solar cycle can shed light on the evolution of the prominences and their FCs. 
However, almost all observational studies focus on one or a few episodes. In contrast, \citet{bashkirtsev_some_1993} studied for the first time the oscillations in a considerable number of filaments. With data obtained from the Sayan observatory between 1981 and 1989, the authors reported 38 oscillations events. They found that the main frequency of oscillations was between 40 and 80 minutes, with a mean value of 60 minutes. In addition, they found a smooth, sinusoidal latitudinal dependence for the observed periods. In our later work \citep{luna_gong_2018}, we did not find a clear relationship between the periods or other properties and the filament latitude. The \citeauthor{bashkirtsev_some_1993} study covered almost a solar cycle, so their latitudinal dependence could be related to the well-known migration of filaments from the poles toward the equator during the cycle.
In \citet{luna_gong_2018} we surveyed prominence oscillations using the GONG network H$\alpha$ data during 2014 January-June, providing an extensive sample of 196 events close to the solar maximum of cycle 24. We found that the oscillations in filaments had an average period of 58 minutes regardless of the amplitude of the oscillation or the type of filament. The oscillations were detected by visual inspection of the data. This visual detection has limitations due to the subjective bias introduced in the study. It is difficult to detect relatively high or small frequency oscillations or periodic motions in small filaments. In addition, a study covering years of observations becomes unfeasible due to a large amount of data to be analyzed by eye.

There are methods to automatically detect and track solar filaments in H$\alpha$ as in \citet{gao_development_2002,shih_automatic_2003,zharkova_seachable_2004,bernasconi_advanced_2005,fuller_filament_2005,bonnin_automation_2013} and \citet{hao_statistical_2015}. These techniques have been used to conduct statistical studies of the properties of filaments such as \citet{jing_relation_2004} that automatically selected a large number of filament disappearances from a data set over 4 years. More recently, \citet{hao_statistical_2015} studied statistically the filaments spanning almost three solar cycles from 1988 to 2013. It is noteworthy that these codes can be used to track the movement of individual filaments. Some of the codes are publicly available such as in \citeauthor{bonnin_automation_2013} We have tried to use this technique to detect and parameterise oscillations without satisfactory results. Although we do not rule out the possibility of using these methods to detect oscillations, we have decided to investigate another approach, which is the one presented in this study. Filament tracking techniques can be useful to extract their characteristics such as length, width, position, etc. also in parallel to the approach followed in this work.

The paper is organized as follows: In \S \ref{sec:data-processing} the automatic technique is described. In \S \ref{sec:synthetic-measurement} we study synthetic data to understand the intensity fluctuations associated with filament oscillations. We also apply the detection method to these data to understand the structures that appear in the spectral analysis. Then in \S \ref{sec:automatic-detection-of-oscillations}, we apply the method to several events where the oscillation has already been reported to test the efficiency of the method. We also see that we apply the method to data where it was not known a priori that oscillations were present. Finally in \S \ref{sec:conclusions} the results are discussed and conclusions are drawn.

\section{Data Processing}\label{sec:data-processing}

GONG's H$\alpha$ data are very useful for detecting oscillations in the filaments. The reason is that the H$\alpha$ images show the dark filaments seen in absorption in sharp contrast to the bright chromosphere around them. In contrast, the SDO EUV images show many highly dynamic structures such as coronal loops that make it difficult to detect filaments and their oscillations. However, GONG data have poorer spatial and temporal resolution and the quality of the data depends on the conditions of the sky at the time of observation. Despite the limitations of the GONG data, the method is robust enough to efficiently detect oscillations in solar filaments as we will demonstrate.

This work aims to explore the possibility of using a technique already applied to detect waves and oscillations from image data cubes \citep[e.g.][]{terradas_two-dimensional_2002,nakariakov_coronal_2007,mcintosh_coherence-based_2008,ireland_automated_2010,auchere_long-period_2014}, in the detection of periodic motions in solar filaments using GONG data.
Following to \citeauthor{auchere_long-period_2014}, the method consists in four steps analysis, namely
\begin{enumerate}
\item Detect a region of interest (ROI) and a temporal sequence. Compensate the solar rotation by remapping to Carrington coordinates by using the \emph{drot\_map.pro solarsoft} routine.
\item Calculate the power spectral density (PSD) for each pixel of the ROI.
\item Identification of the regions with oscillations according to pre-defined criteria.
\item Parametrization and storage of the resulting analysis.
\end{enumerate}
With this methodology the authors detected regions of the solar corona oscillating coherently. In these regions, the intensity fluctuations are very clear with an almost sinusoidal shape. The reason is that these fluctuations are probably associated with changes in the density and temperature of the coronal loops on the LOS of each pixel. In this sense, these fluctuations are intrinsic variations of the emissivity in that portion of the corona. However, the filament oscillations are mainly not intrinsic variations of the intensity of the filament: it is a displacement of part of the dark structure over the solar disk. It is not trivial how the advection of the filament produces the fluctuations of the H$\alpha$ intensity and for this reason, we have included the following section \S\ref{sec:synthetic-measurement}. In that section, we mimic the oscillation of a filament and the spectral analysis of the intensity fluctuations.

For step 2, we analyse the intensity fluctuations for each heliographic pixel of the ROI. We construct the PSD using the fast Fourier transform (FFT) of the light curve. In each pixel $(i,j)$, we define the light curve signal
\begin{equation}\label{eq:inte-normalized}
I_{\mathrm{signal,} i j}=I_{i j}-\bar{I}_{i j} \, ,
\end{equation}
where $I_{i j}$ is the original intensity on each pixel and $\bar{I}_{i j}$ is its time average during the temporal sequence studied. The time signals given by Eq. \eqref{eq:inte-normalized} are additionally normalized to their standard deviation.
The resulting normalized signal is apodized using a Hann window to minimize the effects of the finite duration of the signal on the PSD. We did not apply any detrending nor time-differencing to the signal since these operations are prone to introduce detection artefacts if combined with an improper model of the spectrum of background noise \citep{auchere_fourier_2016}. 
The temporal cadence provided by GONG is of 1 minute which corresponds to a Nyquist frequency of 8.3 mHz. The spectral resolution will depend on the length of the temporal sequence.
In this paper, we have selected time sequences with almost no gaps in the data. When there is a gap, the data is considered zero. We have also used the generalized Lomb and Scargle periodogram from \citet{zechmeister_generalised_2009} and the time-averaged wavelet spectra from the well \citet{torrence_practical_1998}. This allowed us to compare the benefits and drawbacks of each method. The periodogram is very useful when considering time sequences with significant gaps and the wavelets are very useful when there are events where the periodicity changes over time. All three techniques produce identical results for the events considered.
Due to its simplicity we have adopted the FFT for this first study of the feasibility of the oscillation detection technique. In addition, the FFT calculation is faster than the periodogram or the wavelet.
%

In step 3, proper estimation of the expected background power, $\sigma_{i j}$ at each frequency and each pixel is necessary to derive confidence levels. 
For solar coronal time series, the PSD is often of the form of two components
\begin{equation}\label{eq:noise}
\sigma_{i j} = a_{i j} \nu^{-\alpha_{i j}} + b_{i j} \, .
\end{equation}
The first corresponds to the red noise which, as we shall see, is predominant for $\nu<1$ mHz. The second component, the constant $b_{i j}$, corresponds to white noise which is predominant for $\nu\ge1$ mHz. Most of the oscillations in filaments we detect have frequencies below 1 mHz. We also see that the events with frequencies higher than 1 mHz correspond to very small regions not related to filaments, which leads us to conclude that they may be due to false positives or actual oscillations in the solar chromosphere. To make the search more selective we will restrict ourselves to frequencies lower than 1 mHz. However, we fit the function \eqref{eq:noise} \citet{auchere_long-period_2014} to the whole frequency range given by the FFT for correct modeling of the background noise.
To perform this fit of the PSD in each pixel we use the subroutine \emph{curvefit.pro} from IDL
\footnote{Details of this subroutine can be found in \url{https://www.l3harrisgeospatial.com/docs/curvefit.html}. This routine computes the non-linear least squares fit of function \eqref{eq:noise}. It is calculated iteratively until the chi-square varies by a specified amount or the maximum number of iterations is reached. These are given by the parameters TOL and ITMAX which we set to $10^{-4}$ and 1000 respectively.}. 
In this fit, the data are weighted with the inverse of the square of their standard deviation for each frequency. The standard deviation of the PSD is thus equal to its mean at each frequency \citep{auchere_fourier_2016}. However, neither the standard deviation nor the mean value of the PSD is known. For this reason, we use the inverse of the square of the smoothed PSD with a boxcar average of 5 points as a weight in the fit.
In all the cases considered in this work, we assume a global confidence level of $95\%$.
In order to have a positive detection the PSD should be larger than $m \, \sigma_{i j}$ where $m$ is given by
\begin{equation}\label{eq:m-confidence}
m=-\log\left(1 - (1-0.95)^{2/N} \right) \, ,
\end{equation}
being $N$ is the data points in the temporal sequence.
As \citet{auchere_long-period_2014} points out, since a large number of points in the ROI are being analyzed, the probability of having at least one random point above the confidence level is practically one in the whole region. 
However, the probability that these random points appear clustered in adjacent pixels above the threshold decreases rapidly with cluster size. In the study, we analyzed how many pixels appear clustered in adjacent pixels for each frequency and discarded clusters when the number of points is less than $\clustersize$.
We have not found a priori way to fix the $\clustersize$ value, but in events, with clear oscillation in filaments, there are several tens of adjacent points oscillating coherently. In future work to massively automate the detection of oscillations, it will be necessary to provide a criterion for setting $\clustersize$. 
To improve this criterion, we could study which clusters oscillate coherently considering the relative phase of adjacent pixels as in \citet{mcintosh_coherence-based_2008}.
%

GONG data may have oscillations due to artefacts such as atmospheric turbulence, clouds passing in front of the telescope or telescope vibrations. These artefacts can produce quasi-periodic variations in intensity over the entire solar disk. 
This would result in a spike in the PSD that the method could identify as a real oscillation.
To identify these possible artefacts, we calculate the average PSD $\avesigma$. It is computed by averaging the PSD in each frequency over all the ROI pixels. If there is an actual oscillation in filament within the ROI, there will not be a pronounced peak in $\avesigma$. This is because the region that oscillates is small in comparison to the ROI and has a small weight on the average. However, an artefact will produce fluctuations in a large part of the ROI and produce a clear peak in $\avesigma$. In this way by monitoring $\avesigma$ we can detect false oscillations associated with artefacts.
 
\section{Synthetic measurement of prominence oscillations}\label{sec:synthetic-measurement}
\begin{figure}
	\centering\includegraphics[width=\width\textwidth]{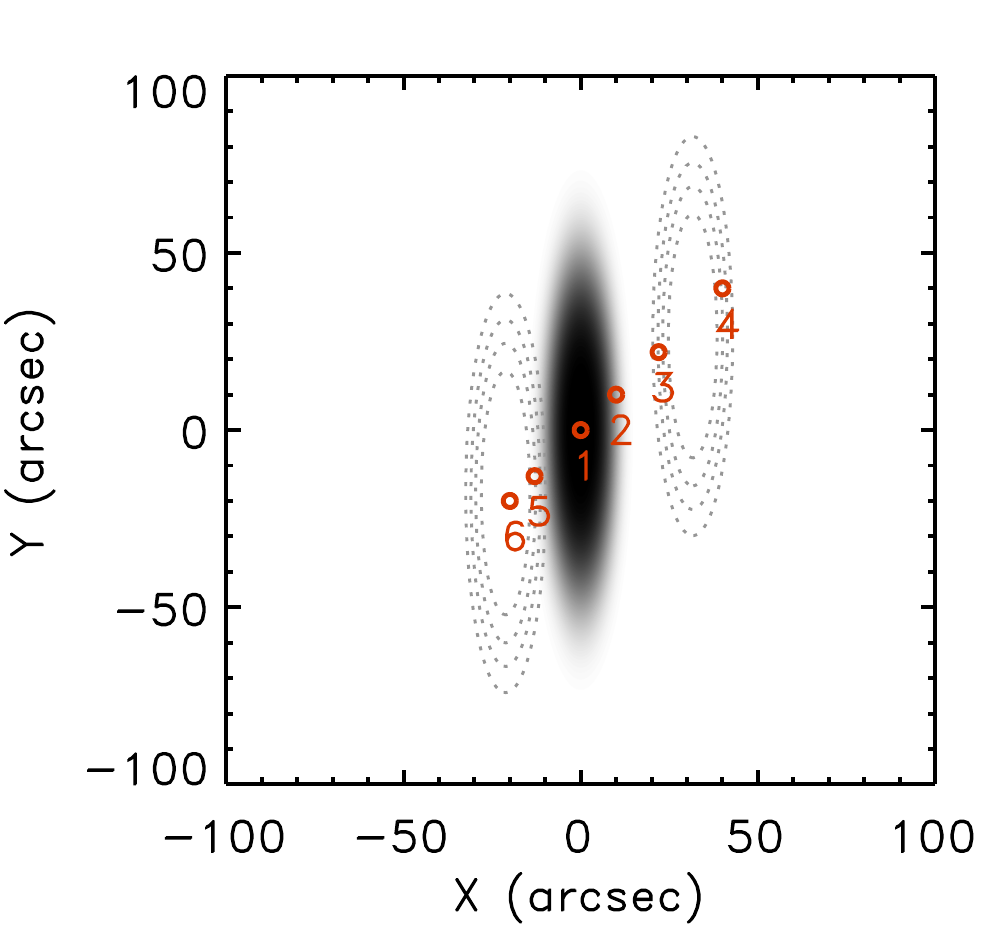}
	\caption{Plot of the intensity given by $I(x,y)$ (see text) mimicking a GONG H$\alpha$ map. The dark area corresponds to the filament seen in absorption. The filament oscillates with a period, $P=60\mins$, damping time, $\tau=75\mins$, and a direction of $45^\circ$ with respect to the $x$-axis. The dotted isocontours at $x > 0$ show the position of the filament at $t=P/4$ when the filament reaches its maximum elongation at the right of the equilibrium position. Similarly the dotted isocontours at $x < 0$ show the filament at $t=3/4 P$ when the filament reaches the maximum elongation at the left. The six red circles show the positions along the oscillations trajectory where the intensity fluctuations are studied in detail in Fig. \ref{fig:single-intensity-fluctuations}. \label{fig:synthetic-filament}}
\end{figure}
\begin{figure*}[!h]
	\centering\includegraphics[width=0.8\textwidth]{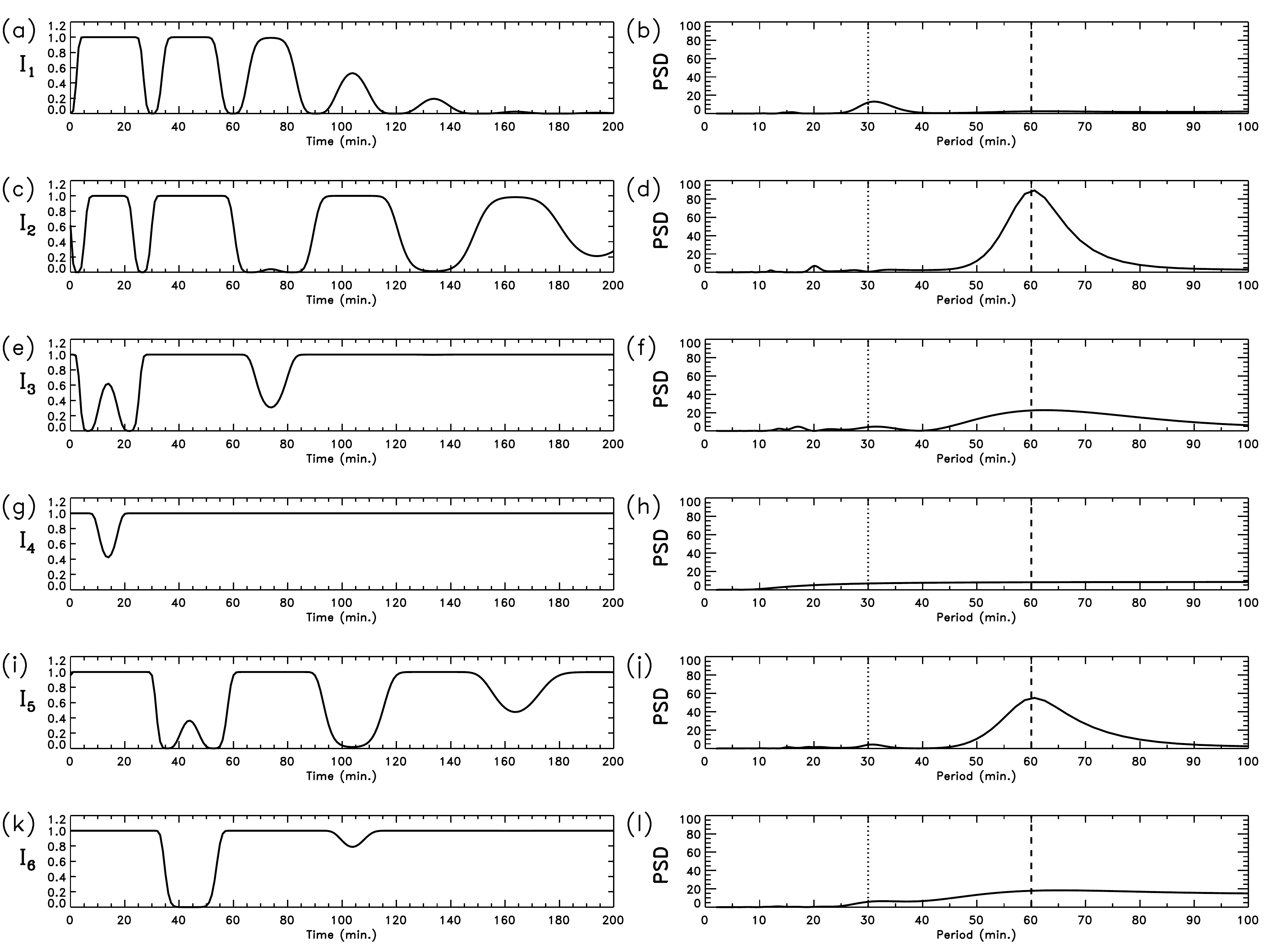}
	\caption{Plot of the intensity fluctuations in the positions (1) to (6) from Fig. \ref{fig:synthetic-filament} in the left panel from top to bottom respectively. The right panel shows the PSD of the corresponding intensity fluctuation. 
	 \label{fig:single-intensity-fluctuations}}
\end{figure*}

The H$\alpha$ images show the filaments in absorption in contrast with the bright chromosphere. 
Oscillations are periodic movements of these dark structures on the bright chromosphere. This motion results in the occultation of the light from the chromosphere around the filament. This occultation produces quasi-periodic fluctuations of the intensity in the areas where the filament moves. However, these fluctuations are not clear sinusoidal patterns as we will see below.
To understand the filament oscillations in the H$\alpha$ images, we reproduce schematically a filament seen in absorption. Fig. \ref{fig:synthetic-filament} shows the H$\alpha$ intensity $I(x,y)$ where the dark structure represents the filament.
This intensity is given as $I(x,y)=I_0-I_\mathrm{Gauss}(x,y)$. The background or chromospheric intensity is $I_0$ and $I_\mathrm{Gauss}$ is a 2D Gaussian with an elliptical shape. For simplicity, we assume that $I_{0}=1$ and the maximum of $I_\mathrm{Gauss}$ is also 1. We reproduce the oscillation by displacing the $I_\mathrm{Gauss}$ along a line that forms $45^\circ$ to the $x$-axis. The period and damping time of the oscillation is set to $P=60$ and $\tau=75\mins$ respectively, which are values in agreement with the observations. The amplitude is set to $A=42$ arcsecs. The maximum elongation of the filament is shown as dotted isocontours at the right of the filament at $t=P/4$. Also, the dotted isocontours at the left correspond to the maximum elongation but at $t=3/4 P$.
In Fig. \ref{fig:synthetic-filament} we have marked six positions where we have probed the intensity fluctuations. These positions are along the trajectory of the periodic motion of the filament. 
The intensity on each point is shown in the left column of Fig. \ref{fig:single-intensity-fluctuations}. In the right column, we have plotted the PSD computed as described in \S\ref{sec:data-processing}.
From the intensity fluctuations (left column) from the selected positions, we see that they do not correspond to clear sinusoidal oscillations. Point 1 corresponds to the centre of the filament (Fig. \ref{fig:synthetic-filament}). In Fig. \ref{fig:single-intensity-fluctuations}(a) we see that the intensity fluctuates between 0 when the filament is at the equilibrium position to 1 when the filament is far from this position. During an oscillation cycle, the filament crosses the central position two times and then the intensity fluctuates two times per period. Fig. \ref{fig:single-intensity-fluctuations}(b) shows the corresponding PSD. The dashed vertical line shows the period of the oscillation of 60 minutes. The dotted vertical lines are half of the oscillation period, 30 minutes. The PSD has a clear peak at around 30 minutes with a maximum value of 13 that is a 20\% of the highest PSD value at position 2. Fig. \ref{fig:single-intensity-fluctuations}(c) shows the intensity at position 2 which is placed around the equilibrium position of the filament at a distance from the filament centre of 14 arcsecs. Panel (d) shows that the PSD has a clear peak centred at 60 minutes with a value of 89 that is the highest value measured at the six positions.
Position 3 is located at 31 arcsecs from the centre and the filament passes over it at around $t=7\mins$, before reaching its maximum elongation on the right side and then when the filament returns to its equilibrium position at $t=22\mins$ approximately. This passage over position 3 is seen as a W shape in the intensity fluctuation of Fig. \ref{fig:single-intensity-fluctuations}(e). Around $t=75\mins$ the filament approaches again to position 3 but due to the damping of the oscillation the filament centre does not reach this position and the intensity fluctuations are reduced. Due to the damping of the oscillation, the filament does not pass this point again in the rest of the time sequence. The PSD shows a very wide peak around 65 minutes with a power of 23 that is a 30\% of the maximum PSD value (Fig. \ref{fig:single-intensity-fluctuations}(f)). In the position 4, at a distance of 56 arcsecs from the centre, the intensity only fluctuates one time (see Fig. \ref{fig:single-intensity-fluctuations}(g)) during the first oscillation cycle. The PSD is not showing a clear peak (Fig. \ref{fig:single-intensity-fluctuations}(h)). In the location 5 (Fig. \ref{fig:single-intensity-fluctuations}(i)) that is located at 18 arcsecs from the centre, the intensity evolution is similar to the case in position 2. The PSD shows a clear peak (Fig. \ref{fig:single-intensity-fluctuations}(j)) with a value of 55 that is a 60\% of the maximum power. 
Position 6 is located at 28 arcsecs from the center that is the maximum elongation at the left of the equilibrium position. The intensity fluctuation in 6 (Fig. \ref{fig:single-intensity-fluctuations}(k)) is similar to the case of position 4 with a no clear peak in its PSD (Fig. \ref{fig:single-intensity-fluctuations}(l)). 
In general with Fig. \ref{fig:single-intensity-fluctuations} we see that even though the intensity fluctuations are not showing clear harmonic oscillations, the PSD may show clear peaks.

\begin{figure}[!h]
	\centering\includegraphics[width=\width\textwidth]{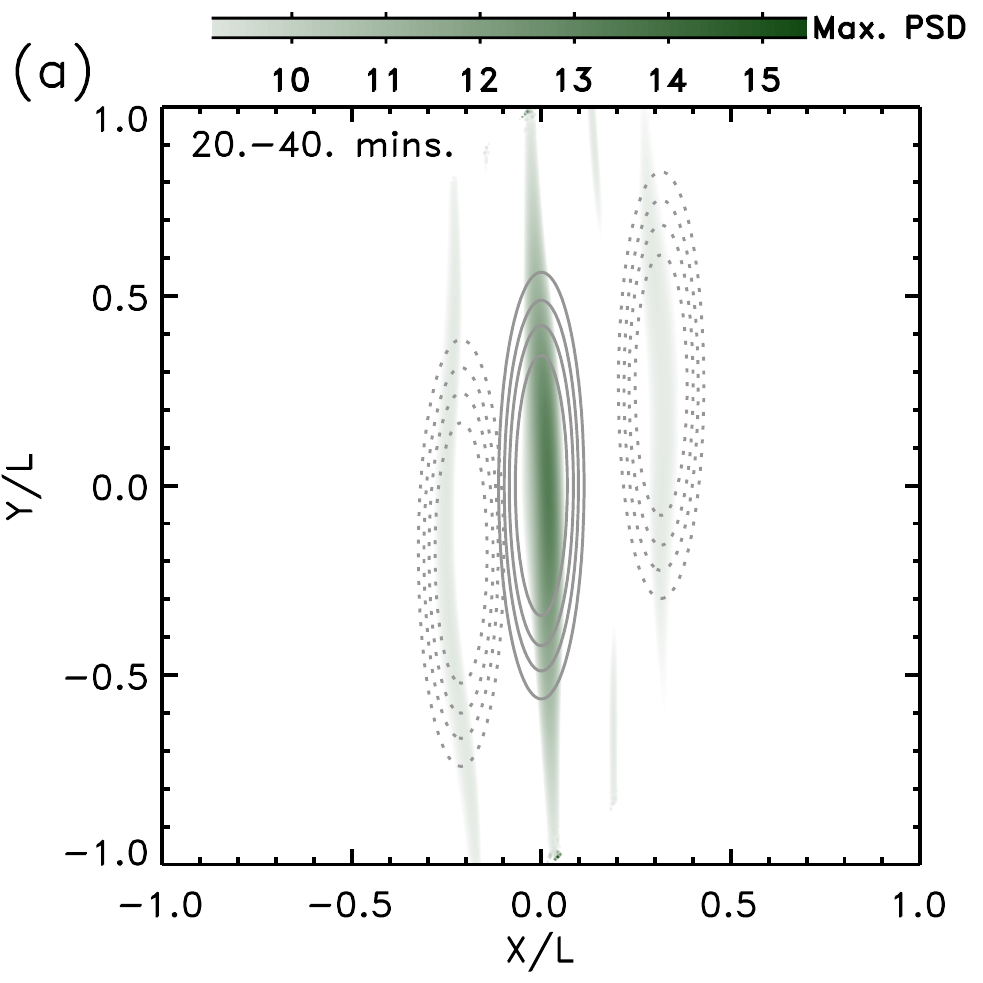}\\
	\centering\includegraphics[width=\width\textwidth]{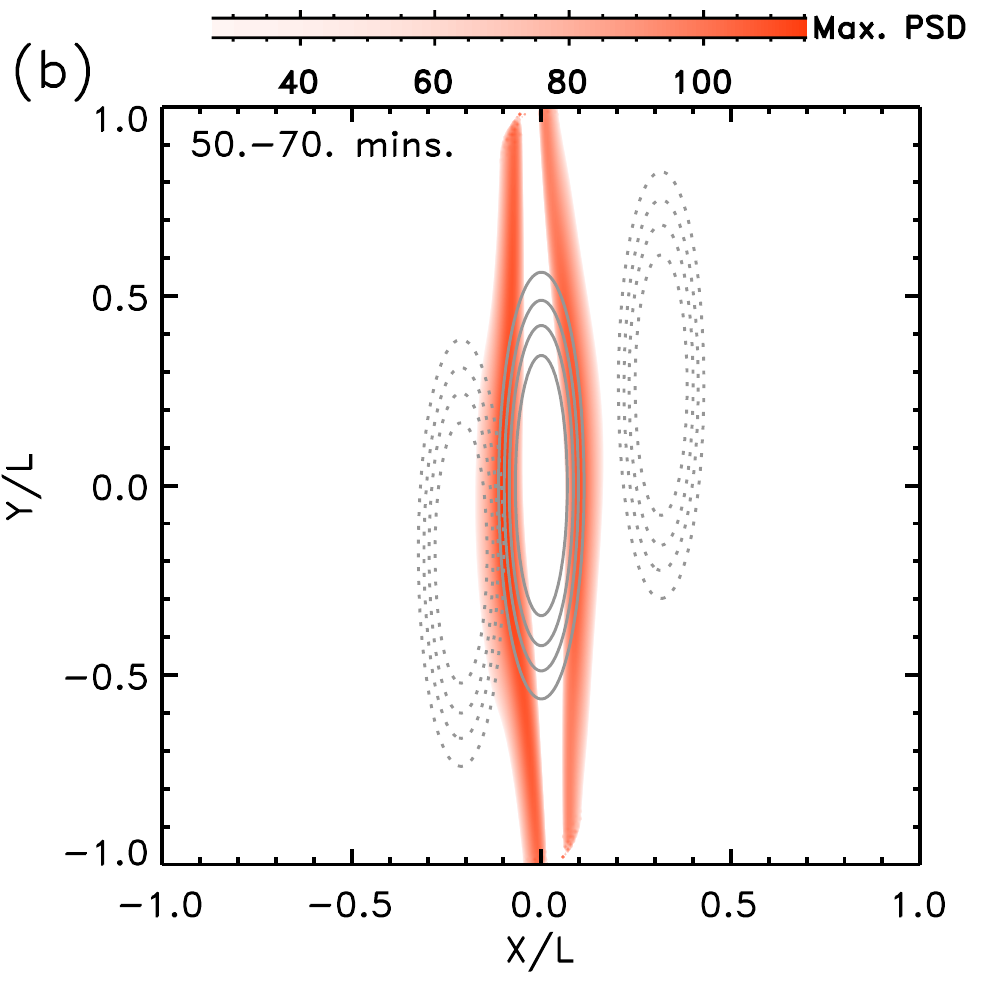}
	\caption{Plots of the PSD projections for (a) the period range 20 to 40 minutes and (b) the period range 50 to 70 minutes. See the text for an explanation of how the projection is constructed. The solid isocontours shows the filament at its equilibrium position. As in Fig. \ref{fig:synthetic-filament} dotted iscontours shows the filament at maximum elongations of the filament at right and left of the equilibrium. \label{fig:psd-cuts}}
\end{figure}

We reproduce the analysis of the oscillations described in \S\ref{sec:data-processing} with the synthetic H$\alpha$ data. In step 1 the ROI is the region shown in Fig. \ref{fig:synthetic-filament}. The duration of the event is 200 minutes and the temporal cadence is 1 minute. For step 2 we generate the PSD for every single pixel of the ROI, $psd_{i j}$. 
The result of the analysis is a data cube consisting of the two spatial coordinates and the third dimension is the period.
To visualize the data, we make projections of the PSD data cube in the $xy$-plane. These projections are slices of the data cube over a range of periods. Within this range of periods, we consider the highest value of the PSD for every single position. Thus, we obtain for each ROI position the highest PSD value in the period range.
Panel Fig. \ref{fig:psd-cuts}(a) shows the projection of the PSD in the plane for the range of periods between 20 and 40 minutes. This is centred in the middle of the oscillation period, 30 minutes. As we have seen in the PSD measured in point 1 (Fig. \ref{fig:single-intensity-fluctuations}(b)), this is due to the filament passing twice through the equilibrium position in each oscillation cycle.
Fig. \ref{fig:psd-cuts}(b) shows the PSD projection in the range of 50 to 70 minutes. The PSDs measured in points 2, 3 and 5 (Figs. \ref{fig:single-intensity-fluctuations}(d), (f) and (j)) correspond to this structure.
In panel (a), we can see that the 30-minute periodicity appears mainly in the central part of the structure with a maximum PSD of 15.5. In (b), we can see the two lobes on both sides of the filament associated with the 60 minutes oscillation with a maximum PSD of 115. The 60-minute periodicity has a much larger PSD than the 30-minute periodicity. As we shall see, this will make the PSD associated with half of the oscillation period difficult to detect and in none of the cases we have been able to observe it. 


\section{Solar filament oscillations}\label{sec:automatic-detection-of-oscillations}


In this section, we apply the technique described in \ref{sec:data-processing} to GONG observations. We first analyse several events from the catalogue by \citet{luna_gong_2018}. In this way, we can contrast the results of the new method with those already known from the catalogue. Second, we use the new technique to detect oscillations in one event not yet reported.

\subsection{The reference case}\label{subsec:jan1sr14}
\begin{figure}
\centering\includegraphics[width=0.45\textwidth]{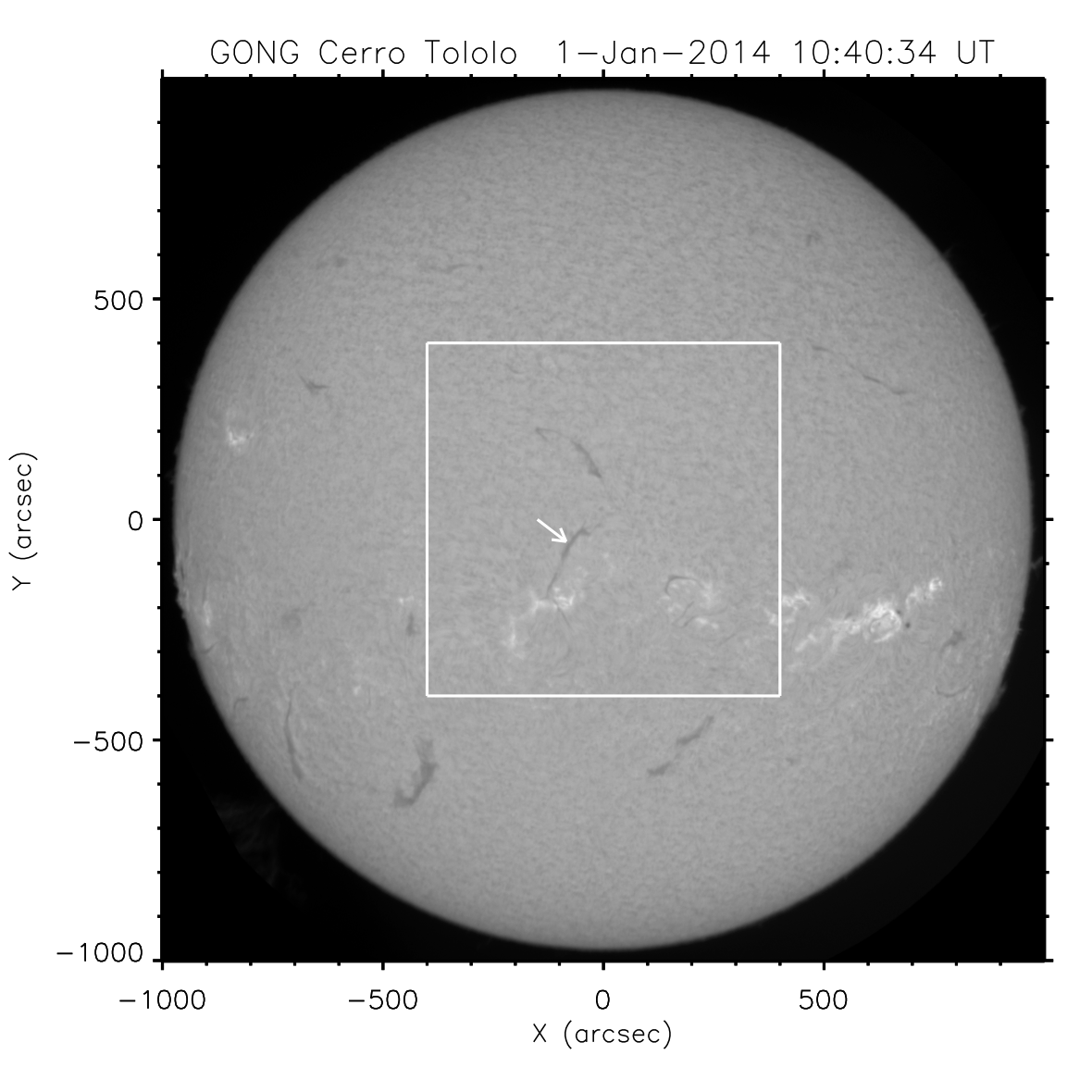}
\caption{Plot showing the first event we are analyzing on January, 1st of 2014. Several filaments are clearly visible as dark structures on the disc. The white box delimits the ROI analysed in the first case considered. The arrow indicates the oscillating filament.\label{fig:first-event-20140101}}
\end{figure}

We start with the first event of the catalogue on January 1st of 2014 as a reference where we describe in detail the methodology used in this research. The period of the oscillation is $76\pm 1 \mins$ with a direction of the plasma forming $16^\circ$ with respect to the filament spine (see details in \citeauthor{luna_gong_2018}'s catalogue). Fig. \ref{fig:first-event-20140101} shows a GONG H$\alpha$ image from the Cerro Tololo telescope of this first event. The white box shows the ROI of $800\times800$ arcs$^2$ we are analysing. Within the ROI there are several filaments and the white arrow indicates the one experiencing the oscillations reported in the catalogue  \citep[solar object locator SOL2014-01-01T07:54:52L178C097 following the IAU convention suggested by ][]{leibacher_solar_2010}. The time sequence analysed lasts 733 minutes from 10:40UT to 22:53UT with a one minute time cadence.
The GONG data have a spatial resolution of approximately 1 arcsecond. 
However, to limit the volume of data to be analysed, we reduce the size of the images to one third by averaging the intensity in squares of $3\times3$ pixels. We will use this reduction on all data analysed in this paper.

\begin{figure}[!ht]
\centering\includegraphics[width=\width\textwidth]{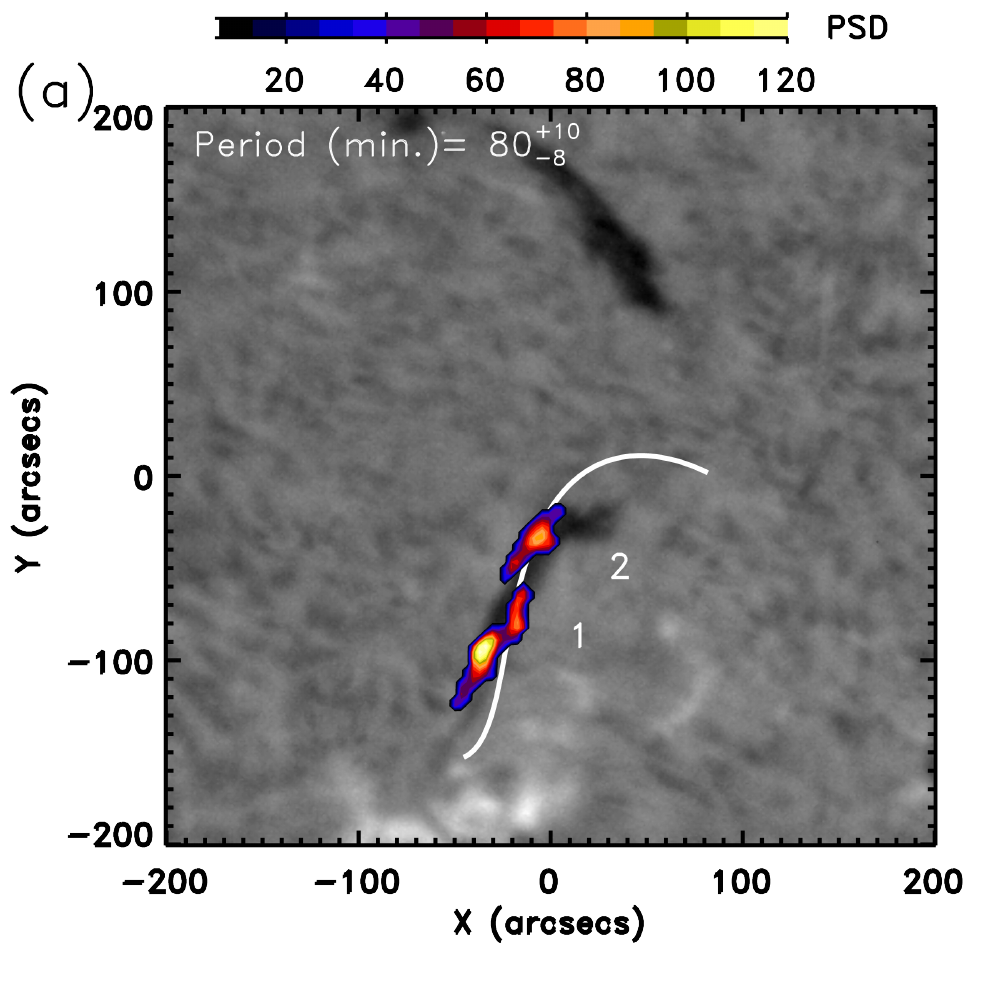}\\
\centering\includegraphics[width=\width\textwidth]{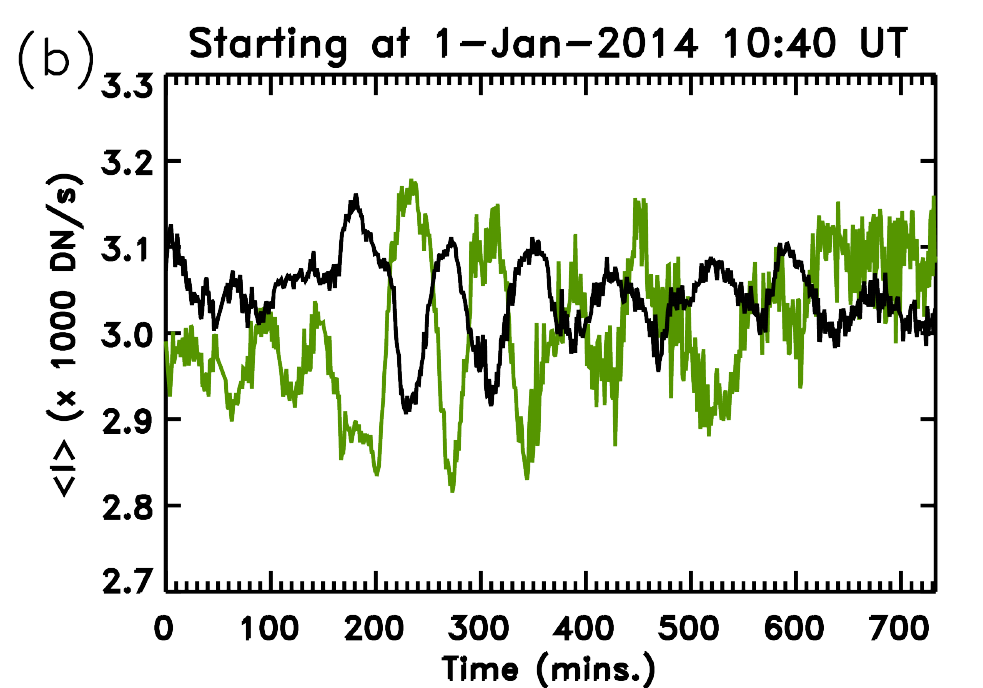}\\
\centering\includegraphics[width=\width\textwidth]{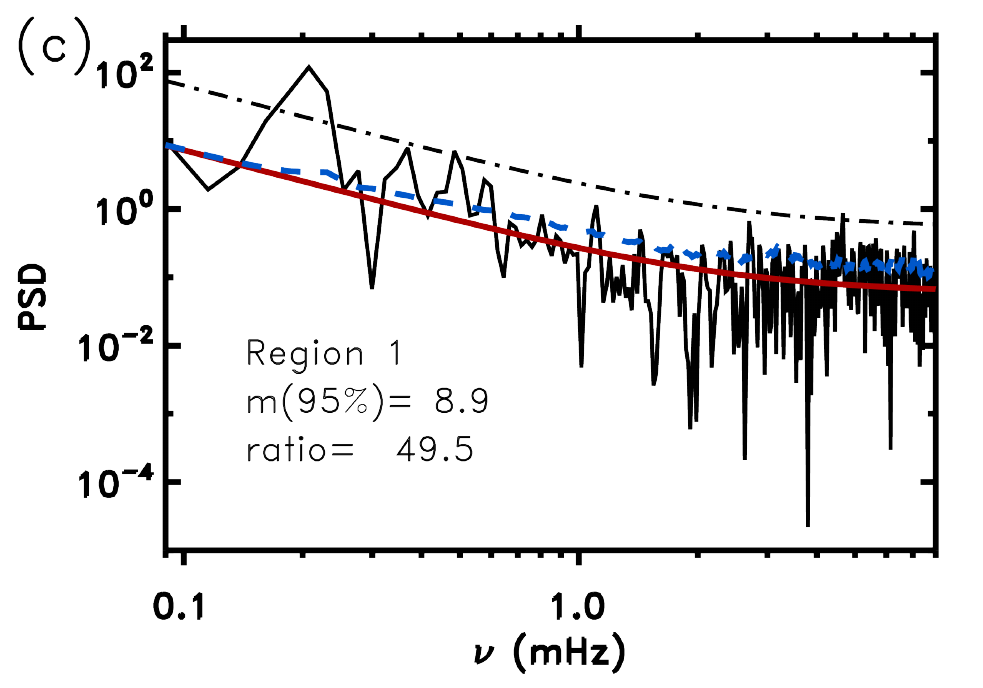}
\caption{Plot showing the results of the automatic detection technique in the event of January 1st, 2014. In (a) the coloured areas are the cut of the whitened PSD (see text). It is plotted over an H$\alpha$ map of the region.
Two clusters have been detected and labelled as 1 and 2. 
The period with maximum PSD and its uncertainty has been printed on the top of the panel.
In (b) the intensity averaged over the whole region 1 (black curve) and the average intensity of region 2 (green curve) are shown. The time is shown on the horizontal axis in minutes starting at 10:40 UT. In (c) the PSD (solid line) with the highest value in the whole ROI is shown that is within region 1. The red lines shows the background noise, $\sigma_{i j}$ and the thick dash-dotted line is the detection threshold, $8.9 \sigma_{i j}$ at the given pixel. The blue dashed line shows the noise averaged over the entire ROI, $\avesigma$. \label{fig:2dmap-detection-labeled-regions-20140101C}}
\end{figure}

In point 2 of the method (\S \ref{sec:data-processing}) we calculate the PSD at each pixel $(i,j)$ of the ROI, $psd_{i j}$. Using Eq. \eqref{eq:noise} we fit the background noise level $\sigma_{i j}$ on each pixel.
In this observation the number of images is $N=723$, then using Eq. \eqref{eq:m-confidence} and considering a confidence level of a 95 \%, $m=8.9$.
We define the whitened data cube as $psd_{i j}(\nu) \ge m \, \sigma_{i j}(\nu)$ and zero when this condition is not fulfilled. The whitened data appears structured into clusters of different sizes. Clusters with sizes below 20 have a very low PSD value and an unclear oscillation pattern. In this event we considered $\clustersize=20$ indicating that clusters with less than 20 points are not considered (see \S \ref{sec:data-processing}). Fig. \ref{fig:2dmap-detection-labeled-regions-20140101C}(a) represents a cut of the PSD data cube in a given period. The period shown is 80 minutes which corresponds to the maximum PSD within the ROI. They are plotted on the H$\alpha$ map of part of the ROI to identify the oscillating structure.  We can distinguish clearly two regions that we have labelled as 1 and 2. 
Region 1 has 90 adjacent points, while region 2 has 65. Region 1 is the largest and also it contains the highest PSD value. 
These two regions are located on both sides of the filament, one at the north of the filament and the other at the south, as we expect from \S \ref{sec:synthetic-measurement}.
The solid white line shows the path used in the catalogue to study the oscillations. This path was constructed by following by eye the motion of the cold plasma during its oscillation. This indicates that the method has detected filament oscillations in the same region as the visual technique of the catalogue.
Fig. \ref{fig:2dmap-detection-labeled-regions-20140101C}(b) shows the average light curve of the pixels within each region. In both regions, the oscillatory pattern is very clear starting at 13:20 UT which is $t=160 \mins$ since the beginning of the temporal sequence. The cool plasma of the filament moves to the North, from regions 1 to 2. Then in region 1, the intensity grows in the first moments of the oscillation whereas in region 2 the intensity decreases because the cool plasma occults the chromospheric emission. The oscillation shows clear signs of damping because the fluctuations decrease with time. The detailed parameterisation of the damping will be the subject of future research.
Fig. \ref{fig:2dmap-detection-labeled-regions-20140101C}(c) shows $psd_{i j}$ (solid line) with largest value shown in Fig. \ref{fig:2dmap-detection-labeled-regions-20140101C}(a). It corresponds to a point inside region 1 at approximately $(x,y)=(-40,-100)\arcsecs$. The red line shows the background noise $\sigma_{i j}$ fitted with Eq. \eqref{eq:noise} function. We see a good fit of the PSD in all the frequencies. For $\nu < 1$ mHz the red noise function fits well the PSD. In this frequency range, the white noise has a negligible contribution. For larger frequencies, the PSD is well fitted by both red and white noise. However, the white noise dominates and the curve becomes flat for increasing frequencies.
The thick dot-dashed line shows the confidence level threshold $8.9 \sigma_{i j}$ computed with Eq. \eqref{eq:m-confidence}. We can see that there is one peak centred at 80 minutes with a power of 49.5$\sigma_{i j}$. This is well above the threshold limit.
In the same figure, we have plotted the averaged PSD of the ROI $\avesigma$ as a blue-dashed line. This is not showing peaks indicating that there are no possible artefacts. 

From this event, we see that the automatic technique works very well. The obtained period of 80 minutes is similar to the 76 minutes from the catalogue. The small discrepancy may be due to the poor spectral resolution of the PSD. We can estimate the error in the period determination given by the spectral resolution. In this case, the spectral resolution is 0.023 mHz and then the period ranges from 72 to 90 minutes or $80^{+10}_{-8}$ minutes. This estimate can be taken as the first approximation of the period uncertainty \citep[see][]{gregory_bayesian_2001,vanderplas_understanding_2018}. Considering the period uncertainty both methods agree well. In addition, the two regions are shown in Fig. \ref{fig:2dmap-detection-labeled-regions-20140101C}(a) correspond to the region where we can visually detect periodic motions in agreement with the catalogue findings. We have not detected the PSD distribution with half of the period centred at the filament as in Fig. \ref{fig:psd-cuts}(a). It is probably below the detection threshold.
This indicates that the intensity fluctuations in individual positions are equivalent to $I_2$, $I_3$ or $I_5$ shown in Figs. \ref{fig:single-intensity-fluctuations}(c), \ref{fig:single-intensity-fluctuations}(e) and \ref{fig:single-intensity-fluctuations}(i) respectively with no presence of half period signal.
%

\subsection{Oscillations in a tenuous quiescent filament near the limb}\label{subsec:feb13th14}

\begin{figure}[!ht]
\centering\includegraphics[width=\width\textwidth]{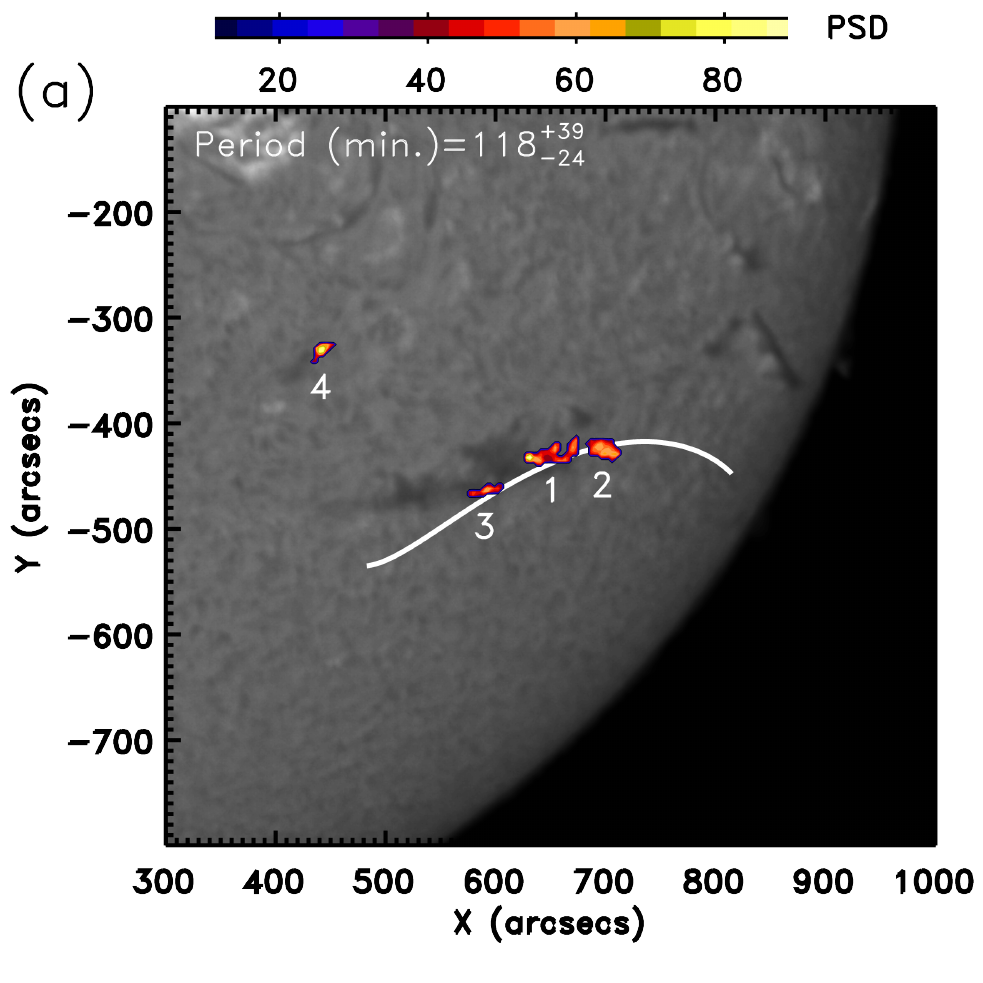}
\centering\includegraphics[width=\width\textwidth]{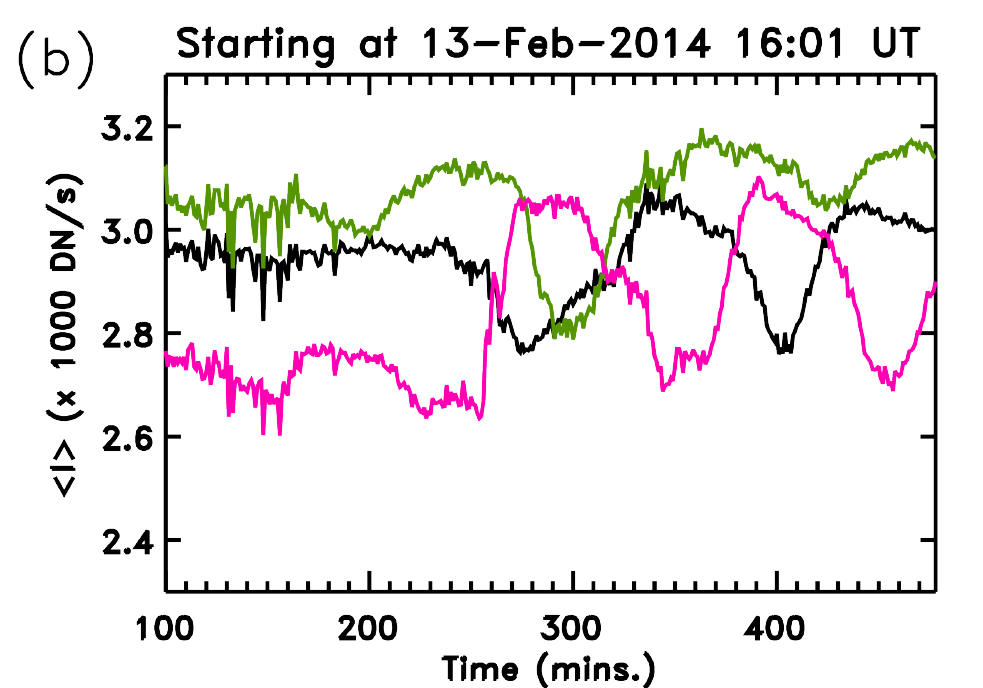}
\centering\includegraphics[width=\width\textwidth]{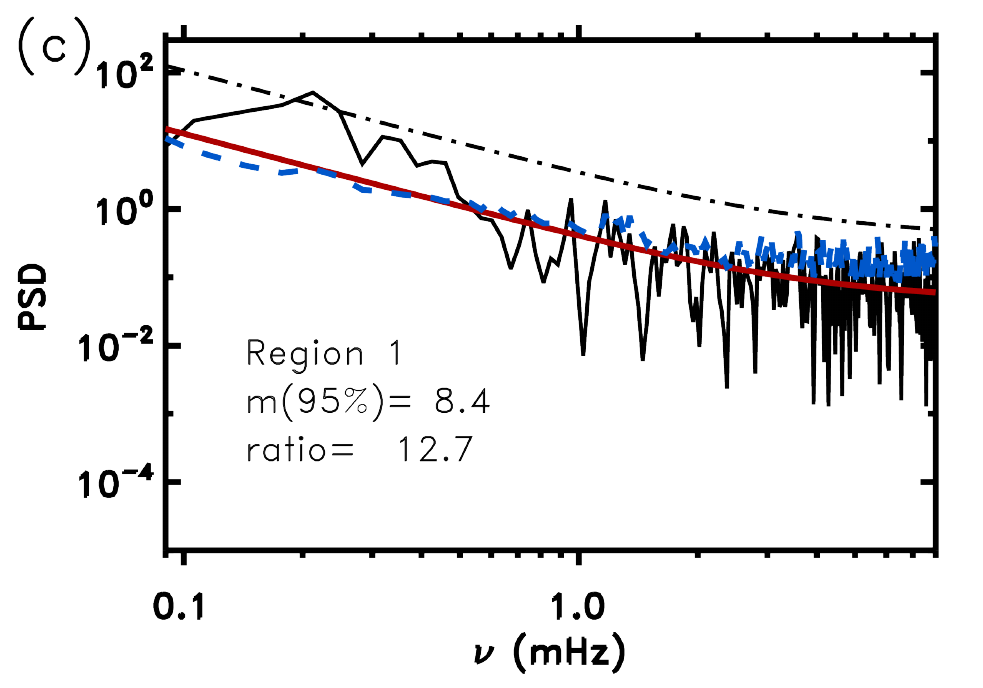}
\caption{Plot showing the event of February 13th, 2014 as in Fig. \ref{fig:2dmap-detection-labeled-regions-20140101C} with similar panels and annotations. We have detected four regions labeled from 1 to 4 in (a). Panel (b) shows the averaged intensity in region 1 (black), 2 (green) and 3 (pink). The maximum PSD is located inside region 1 and shown in (c).\label{fig:2dmap-detection-labeled-regions-20140213B}}
\end{figure}

The second event we consider is event number 63 from the catalogue on February 13rd of 2014 with an oscillation period of $103\pm1 \mins$. It corresponds to a quiescent filament located near the solar limb (SOL2014-02-13T19:31:28L015C123).
It is interesting to test the method with this event because the quiescent filament is very faint. As the filament is close to the limb we also want to check the effect of limb darkening on the detectability of the oscillations.
Fig. \ref{fig:2dmap-detection-labeled-regions-20140213B}(a) shows the ROI we are considering. The technique shows three regions labeled from 1 to 3 with clear oscillations. 
There is a fourth region labelled 4, which corresponds to the oscillation in a small filament. This oscillation in region 4 can be detected in the GONG data by eye. However, in the following, we will focus on the oscillations of the large filament associated with regions 1 to 3.
The three areas are above the quiescent filament where the oscillation was reported in event number 63 of the catalogue.
In addition, we have plotted the path of the slit used in the catalogue and we can see that the three areas are over this path.
The period is around $118^{+39}_{-24} \mins$ which is in agreement with the $103\pm1 \mins$ reported in the catalogue. 
The uncertainty in the period is very large in this event. The reason is that in the range of the period of the oscillation the resolution in the periods is very small. The resolution can be increased by considering larger temporal sequences combining observations from consecutive telescopes of the network. For example, in this observation, the data comes from the Big Bear telescope. We can complete the temporal sequence by using the data from Mauna Loa's telescope or Learmonth depending on the quality of the data. However, this will be the subject of future research when we will analyze massively the GONG data by combining all the telescopes. 
The averaged light curves in all regions are shown in Fig. \ref{fig:2dmap-detection-labeled-regions-20140213B}(b). The oscillations are very clear and the light curves have different phases. 
The oscillations from regions 1 and 3 are completely out of phase. This indicates that both regions are the lobes at both sides of the filament as in the previous case or from Fig. \ref{fig:psd-cuts}(b). Region 2 is delayed 20 minutes with respect to region 1. Region 2 also belongs to the same oscillation and the same lobe as region 1. However, it appears that the oscillating plasma moves behind parts of the filament that do not oscillate, producing this pattern.
The PSD with the largest amplitude corresponds to the region 1 shown in Fig. \ref{fig:2dmap-detection-labeled-regions-20140213B}(c). In this observation we consider we have detection when the PSD is larger than 8.4$\sigma_{i j}$. In this case the PSD peak is 12.7$\sigma_{i j}$ that is above the threshold limit.
Both $\sigma_{i j}$ and $\avesigma$ have similar values in all the frequency domain shown in (c) indicating that there are no global oscillations of the GONG images associated with artefacts (see \S \ref{sec:data-processing}). 

\subsection{Very regular oscillation in an AR filament}\label{subsec:feb9th14}
The third case we analyze is event number 58 in an active region filament (SOL2014-02-10T19:01:37L000C109). In the catalogue, we found a very clear oscillatory pattern with a period of $47 \pm 1 \mins$. In Fig. \ref{fig:2dmap-detection-labeled-regions-20140209C}(a) we see that the automated method has found oscillations in the same filament. We also see that there are no more oscillations in the ROI. As in previous cases, there are two regions on both sides of the filament labelled as 1 and 2 in the figure. The period of the oscillation is $44_{-3}^{+3} \mins$ which is in very good agreement with the catalogue's value.
In the panel, we have also plotted the slit path to construct the time-distance diagram in the catalogue. We see that regions 1 and 2 are not completely over the path. In this case, this discrepancy is because the path selected in the catalogue is not optimal. If we had considered the path over the two regions the oscillation in the catalogue's time-distance diagram would be even clearer.
In panel (b) the averaged intensity in both regions is shown. Both show a very clear oscillatory damped sinusoid pattern. In contrast to previous cases, both oscillations are not completely in antiphase. 
It might be due to that the plasma does not oscillate as a solid body as in \S \ref{sec:synthetic-measurement}. Thus, different parts of the filament could oscillate with different phases.
The periodogram shows an important peak of 114.7$\sigma_{i j}$ at 0.38 mHz, that is much larger than the detectability threshold, $8.8 \sigma_{i j}$.
\begin{figure}[!ht]
\centering\includegraphics[width=\width\textwidth]{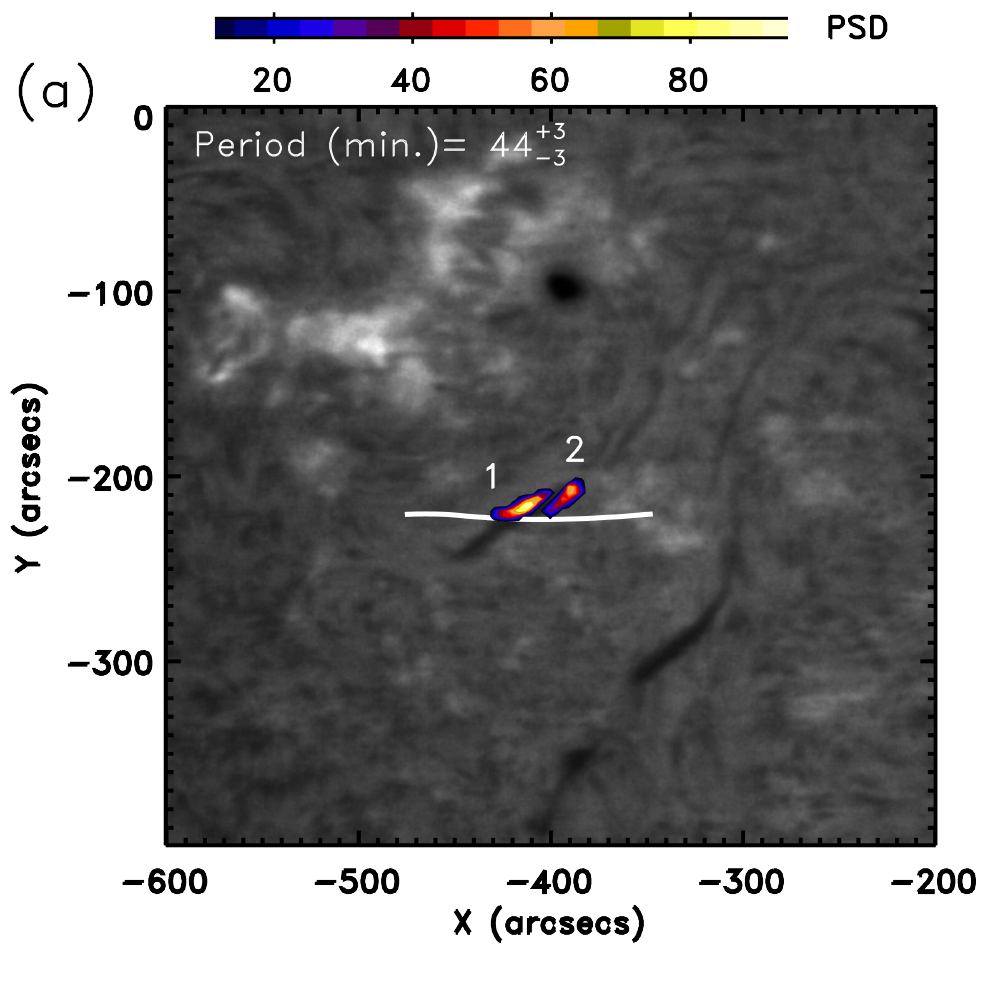}
\centering\includegraphics[width=\width\textwidth]{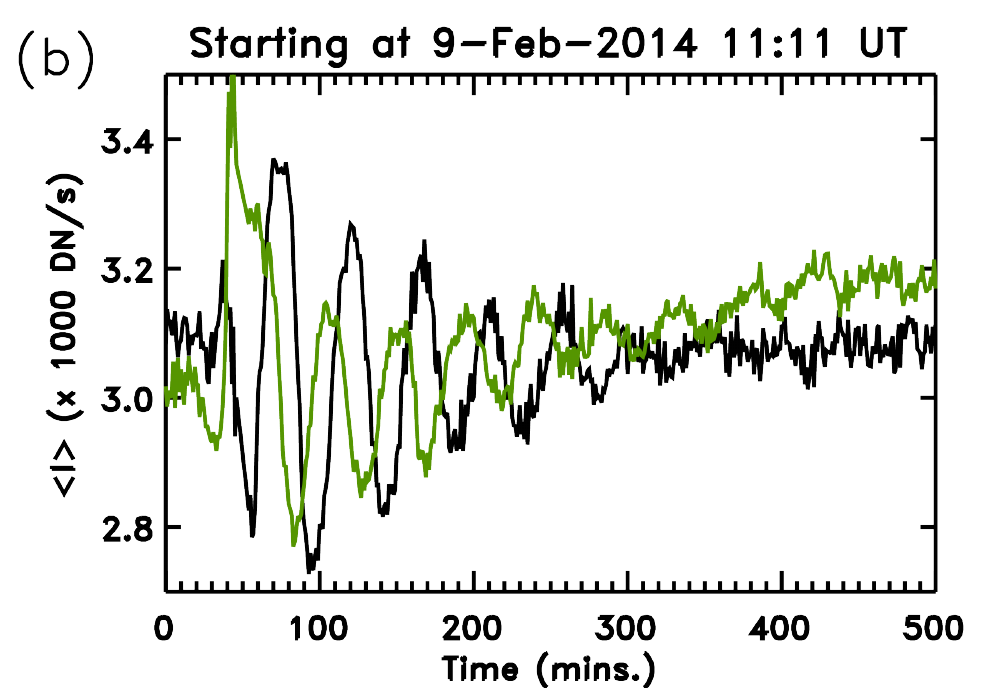}
\centering\includegraphics[width=\width\textwidth]{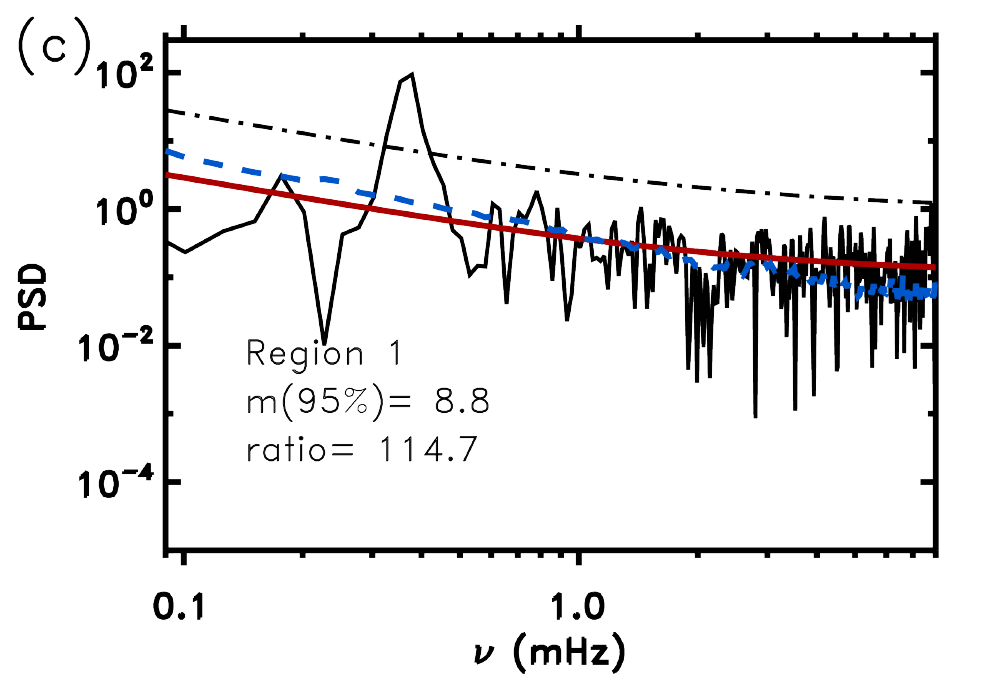}
\caption{Plot showing the event of February 9th, 2014 as in Fig. \ref{fig:2dmap-detection-labeled-regions-20140101C} with similar panels and annotations.\label{fig:2dmap-detection-labeled-regions-20140209C}}
\end{figure}

\subsection{Two extreme cases: events with very small and very large amplitudes}\label{subsec:jun16th14}
In the next two examples, we analyze two extreme cases, one with very small amplitude and the other being a very energetic event. The first event corresponds to the case with one of the smallest velocity amplitudes in the catalogue. It corresponds to a double oscillation with numbers 189 and 190 from the catalogue with periods around 40 minutes and the velocity amplitude is of just fem $\kms$.
\begin{figure}[!ht]
	\centering\includegraphics[width=\width\textwidth]{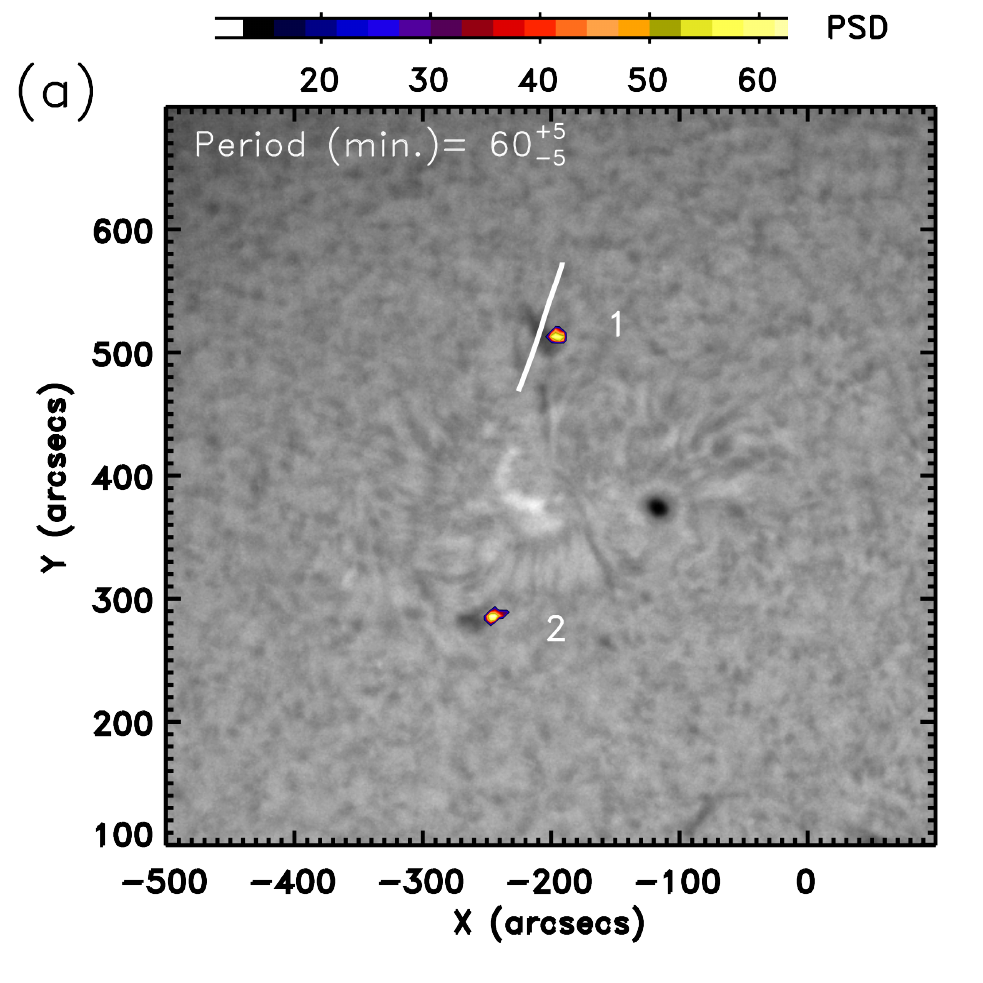}
	\centering\includegraphics[width=\width\textwidth]{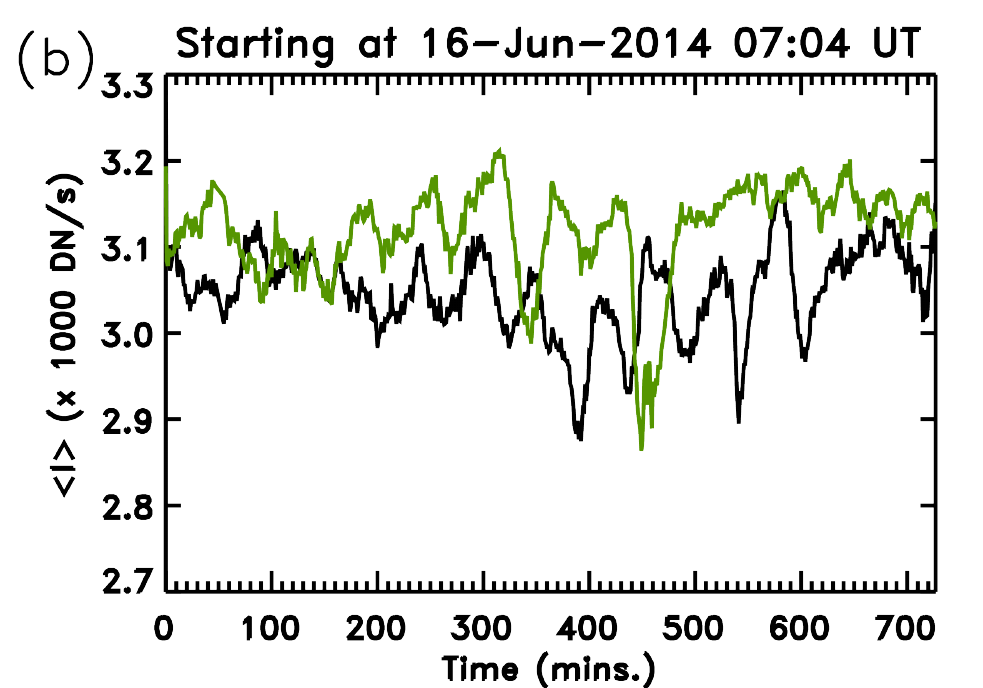}
	\centering\includegraphics[width=\width\textwidth]{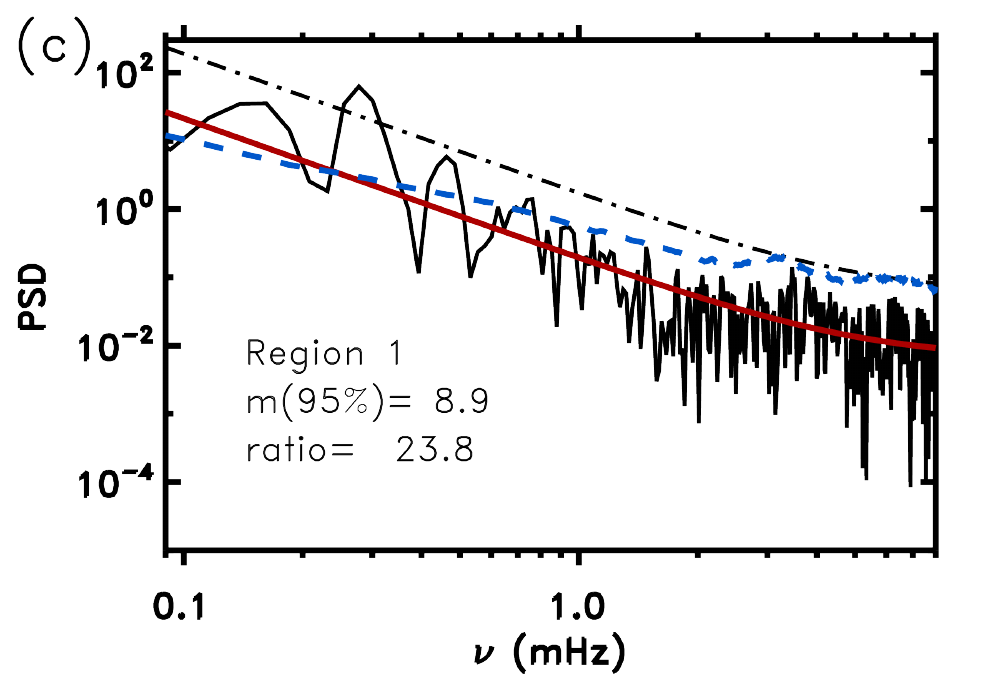}
	\caption{Plot showing the event of June 16th, 2014 as in Fig. \ref{fig:2dmap-detection-labeled-regions-20140101C} with similar panels and annotations.\label{fig:2dmap-detection-labeled-regions-20140209C}}
\end{figure}
Fig. \ref{fig:2dmap-detection-labeled-regions-20140209C}(a) shows the ROI and the regions in color where we have detected oscillations. We have detected two regions where the one labelled 2 has not been reported in the catalogue. In the two regions, the oscillation period is around 60 minutes. Region 1 belongs to the oscillation already reported in the catalogue in the filament (SOL2014-06-16T18:37:17L136C061) in the north of the sunspot. However, we see that the period is quite different from the one already reported. We have checked the data cube containing all frequencies and have not detected a periodicity of 40 minutes. The discrepancy may be because in the time-distance diagram of the catalogue, the oscillation is of small amplitude and there is only one cycle. Under these conditions, the method used in the catalogue probably failed. We have plotted the path used to track the oscillation in the catalogue. We can see that the path and region 1 are close but not exactly in the same part of the filament. This also may also contribute to the discrepancy between both methods. Region 2 belongs to a small filament (SOL2014-06-16T18:37:17L132C071) located south of the sunspot.
In panel (b) the averaged light curves in each of the regions are shown. Periodic fluctuations are clear in both curves. Panel (c) shows that the PSD has a clear peak above the background noise. 
The PSD peak is 23.8$\sigma_{i j}$ which is well above the detection threshold of 8.9$\sigma_{i j}$.
Visual inspection of the GONG data shows the oscillation in the two filaments, which corroborates the efficiency of the method presented here.

\begin{figure}[!ht]
	\centering\includegraphics[width=\width\textwidth]{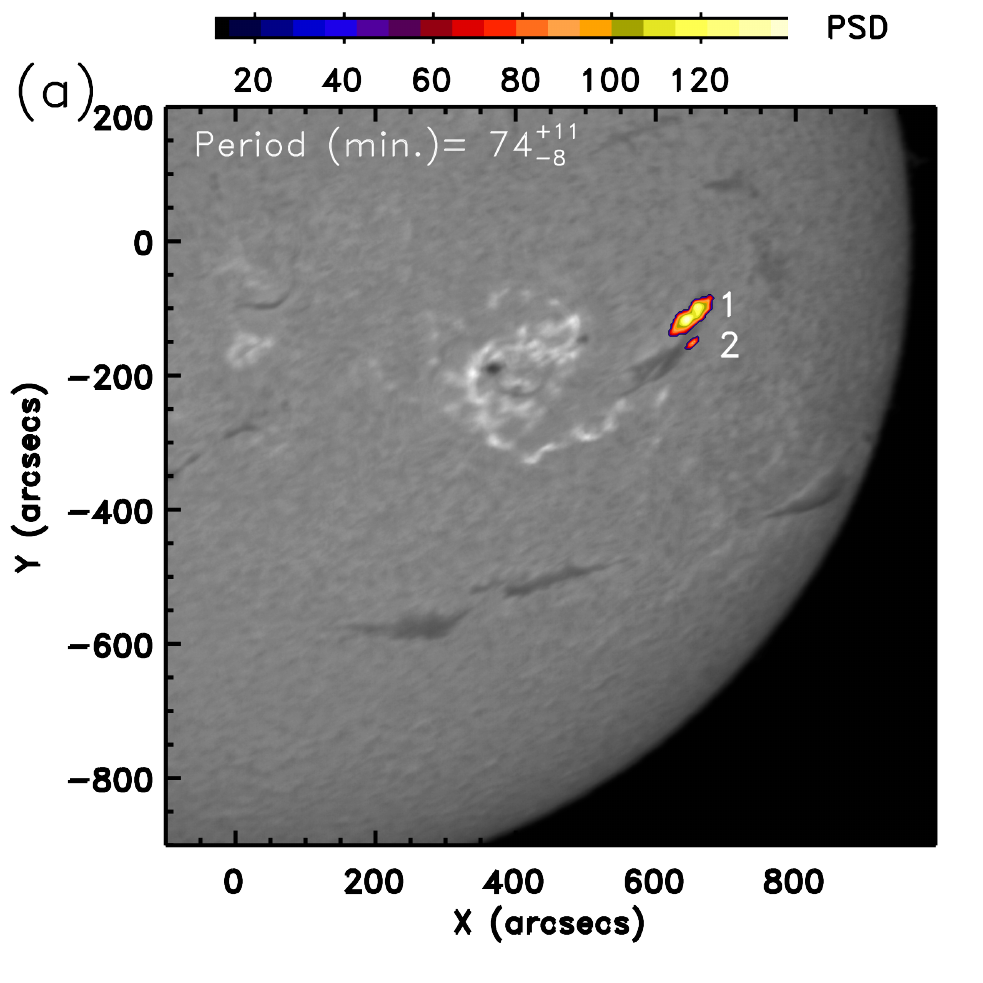}
	\centering\includegraphics[width=\width\textwidth]{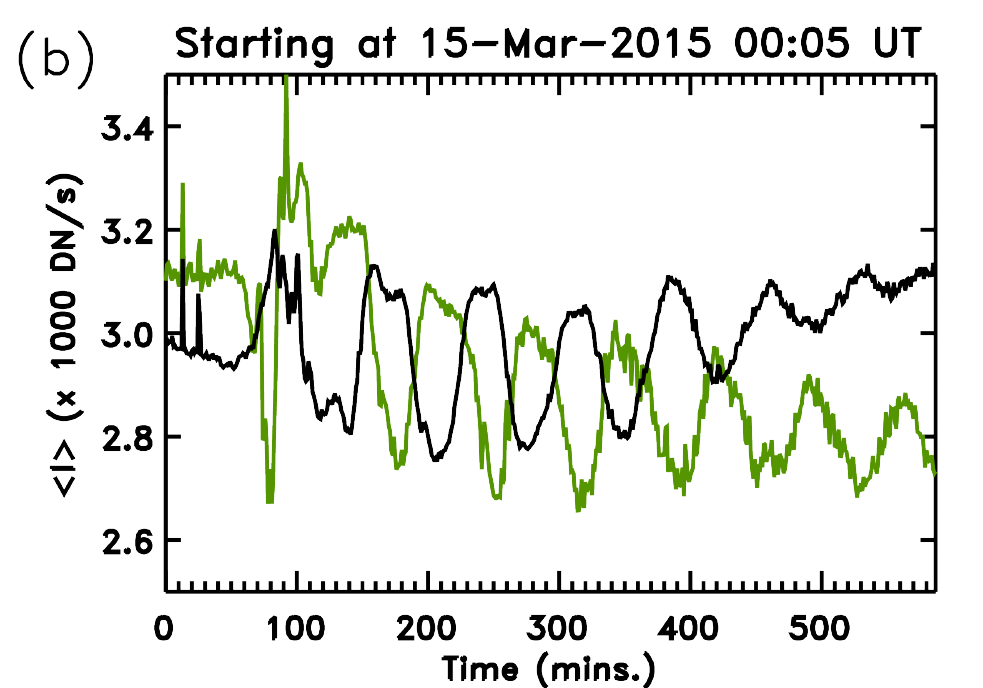}
	\centering\includegraphics[width=\width\textwidth]{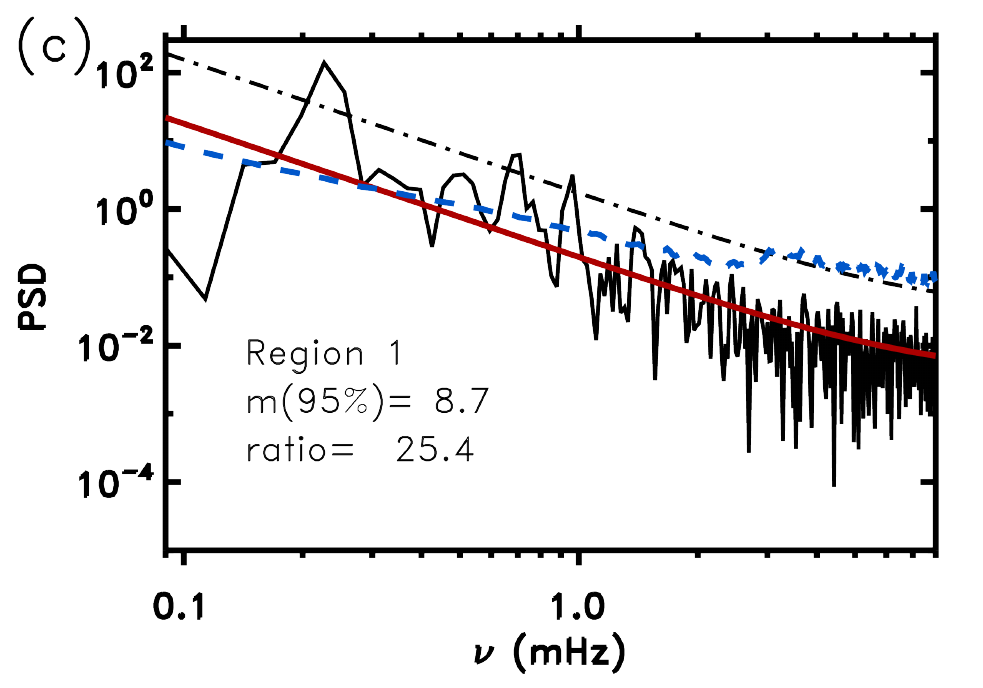}
	\caption{Plot showing the event of March 15th, 2015 as in Fig. \ref{fig:2dmap-detection-labeled-regions-20140101C} with similar panels and annotations.\label{fig:2dmap-detection-labeled-regions-20150315L}}
\end{figure}
The following event corresponds to a very energetic case where its oscillations have not been described in the catalogue or the literature. However, the oscillation is one of the clearest and with one of the largest amplitudes we have seen in the GONG and SDO/AIA data. These oscillations are associated with a flare produced by a two-step eruption on 15 March 2015. The eruption has been reported by \citet{chandra_two-step_2017}. In Fig. \ref{fig:2dmap-detection-labeled-regions-20140209C}(a) we see the ROI and the two regions where the technique has detected the oscillations. The two areas are around the filament (SOL2015-03-15T08:19:18L211C105) and correspond to a $74^{+11}_{-8}$-minute oscillation.
The averaged light curves in the two regions clearly show the periodic fluctuations. 
In panel (c) there is a clear peak in the PSD that is 25.4$\sigma_{i j}$ that is much larger than the detection threshold, $8.4\sigma_{i j}$. 
We can also see two additional peaks at 0.7 mHz and 0.95 mHz above the detection threshold. In this case, the oscillation of the filament is a combination of three different periodicities.

The $\avesigma$ is not showing peaks in the two cases presented in this section (Figs. \ref{fig:2dmap-detection-labeled-regions-20140209C}(c) and \ref{fig:2dmap-detection-labeled-regions-20150315L}(c)). This indicates that there are no possible artefacts as discussed in \S \ref{sec:data-processing}. However, in both cases, $\avesigma$ has a steeper slope than $\sigma_{i j}$ for increasing values of the frequency. In addition, $\avesigma$ increases abruptly for $\nu > 2$ mHz becoming much larger than $\sigma_{i j}$. In the case of June 16th, 2014 we have found that the PSD has an important contribution at high frequencies in the plage region located at the centre of Fig. \ref{fig:2dmap-detection-labeled-regions-20140209C}(a) and also from the spot seen in the same panel. The contribution from the plage and the spot has a strong influence on the average PSD $\avesigma$ over the whole ROI. Similarly for the March 15th, 2015 there is a contribution to the high frequencies from the plage and flaring region seen as bright area centered at $(x,y)=(-250,400)$ arcsecs in Fig. \ref{fig:2dmap-detection-labeled-regions-20150315L}(a).

\subsection{Test case}\label{subsec:feb9th14}
In the latter case, we apply the detection method to a GONG temporal sequence of February 9th, 2022. On that day there were some filaments in the solar disc and it is not known a priori if there are oscillations in them. 
\begin{figure}[!ht]
	\centering\includegraphics[width=\width\textwidth]{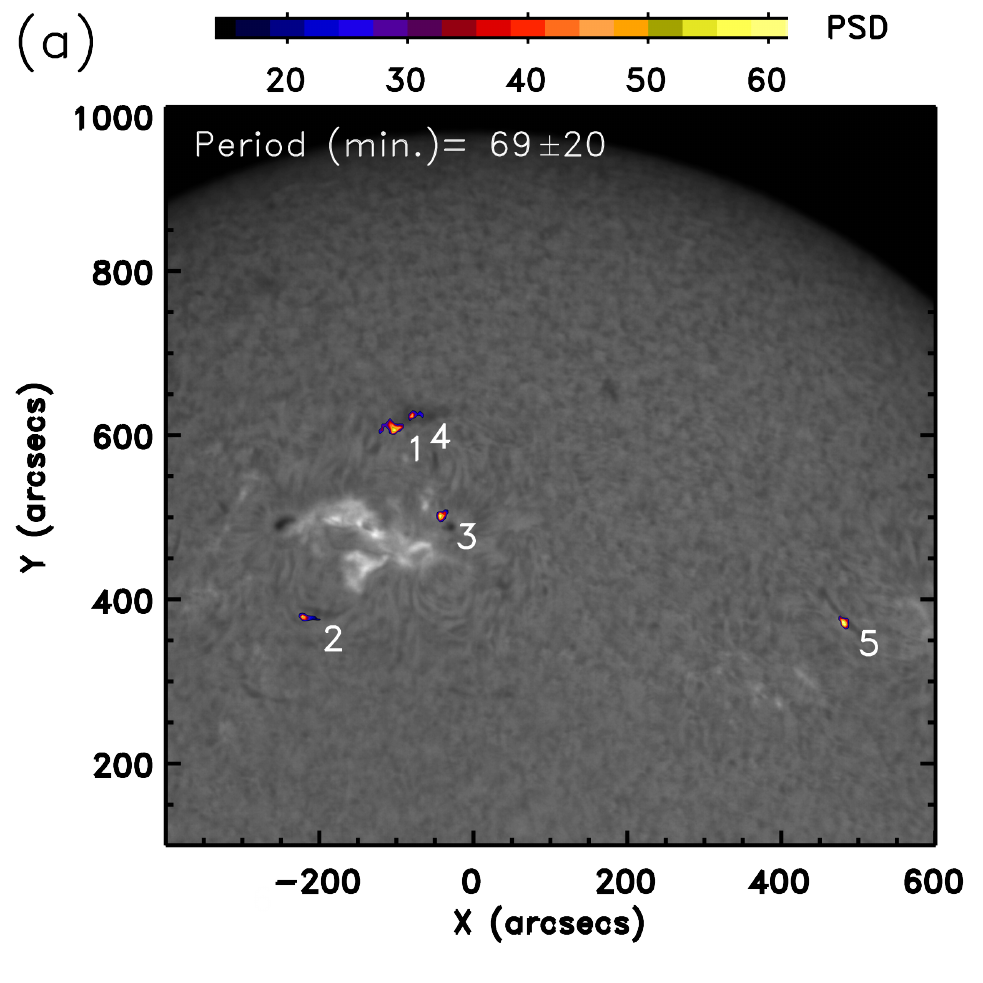}
	\centering\includegraphics[width=\width\textwidth]{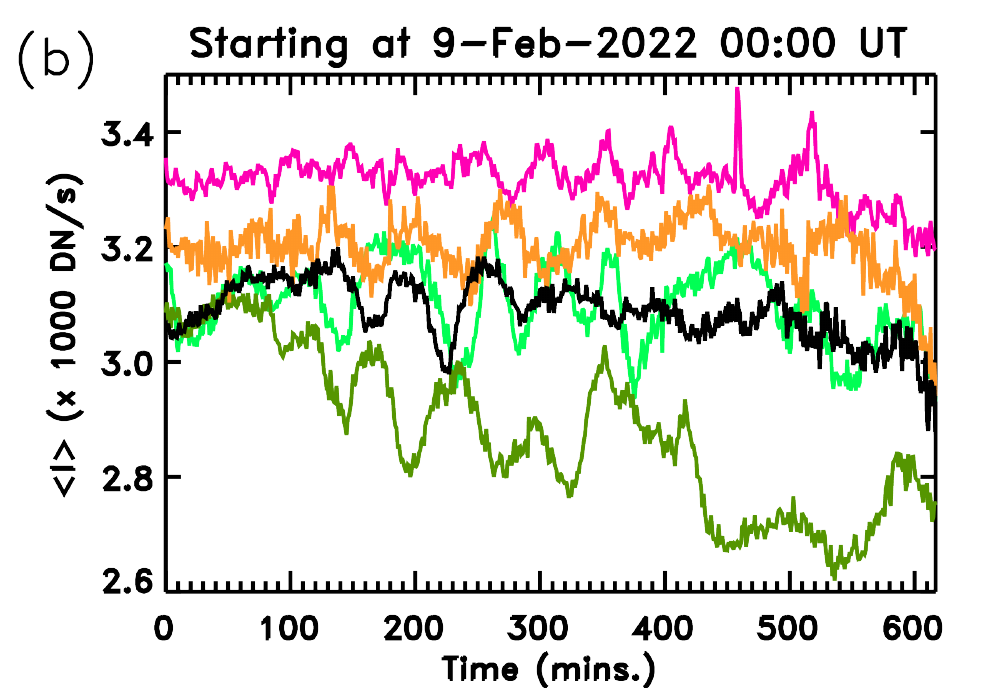}
	\centering\includegraphics[width=\width\textwidth]{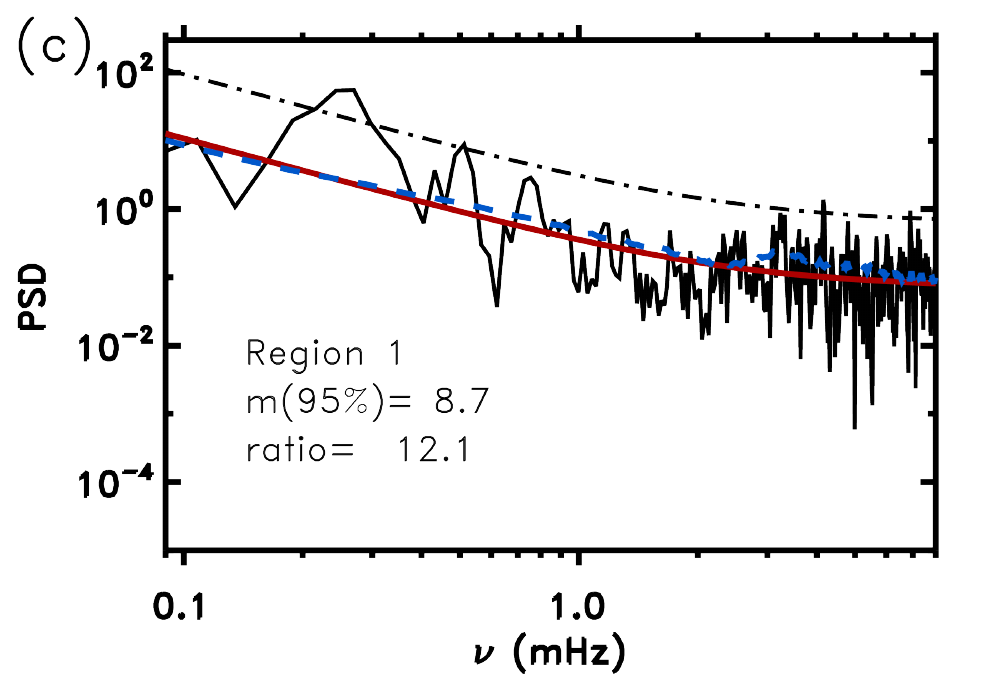}
	\caption{Plot showing the event of February 9th, 2022. In contrast to Fig. \ref{fig:2dmap-detection-labeled-regions-20140101C}, in (a) we show several PSD projections in a given range from 49 to 89 minutes. In panel (b) the averaged intensity in 1 (black), 2 (green), 3 (blue), 4 (orange) and 5 (red) are plotted.
The periods for the regions 1 to 5 are respectively $69^{+9}_{-7}$, $51^{+5}_{-4}$, $51^{+5}_{-4}$, $62^{+7}_{-6}$, and $69^{+9}_{-7}$ minutes.\label{fig:2dmap-detection-labeled-regions-20220209L}}
\end{figure} 
We analyzed the whole disk and we have detected oscillations in five regions. Fig. \ref{fig:2dmap-detection-labeled-regions-20220209L}(a) shows part of the ROI where the oscillations have been detected. The panel is not equivalent to the previous cases where we have plotted a cut of data volume in a given frequency. 
In contrast, we make a projection of the PSD in the ROI similar to Fig. \ref{fig:psd-cuts}. The range of periods considered is from 49 to 89 minutes. Within this period range, we obtain the highest PSD value for every position of the $xy$-plane.
%
Regions 1 and 4 are over the same filament (SOL2022-02-09T02:26:30L340C057) located at the north of the active region, AR12941. Region 2 is associated with the oscillations in a filament (SOL2022-02-09T02:26:30L334C074) located at the south of the same AR. In the three regions, we can identify very well the oscillations by eye from the GONG data and the intensity fluctuations are clear in panel (b). The largest PSD is inside region 1 and it is plotted in panel (c). It is 12.1$\sigma_{i j}$ that is above the 8.9$\sigma_{i j}$ of the detection threshold.
Regarding region 3 the oscillation is less clear. A visual inspection of the H$\alpha$ data shows a small dark area that seems to be associated with repetitive surges just north of the sunspot. In panel (b, pink curve) we can see those quasi-periodic fluctuations of the intensity in the region. It seems that the intensity fluctuations increases in amplitude from 100 to 520 minutes after the initial time and the fluctuations cease shortly after. 
Finally, the oscillation in region 5 corresponds to the periodic motion in an intermediate filament 
(SOL2022-02-09T02:26:30L018C074) located between AR12940 (not shown) and AR12941. In this case, the oscillation is also visually identified in the data although it is not as clear as in 1 and 2. Panel (b) shows the periodic fluctuation of the intensity in the region.
This case shows that the technique can detect true oscillations not easily detectable by eye but also quasi-periodic fluctuations probably not associated with actual oscillations. In the future, we should define criteria to distinguish real prominence oscillations or quasi-periodic fluctuations.

\section{Discussion and Conclusions}\label{sec:conclusions}
This work explores the possibility of using a spectral technique to detect automatically oscillations in solar filaments.  This studies the periodic fluctuations of the intensity in each pixel of the considered data. For each pixel, the PSD is constructed using the FFT.  Detection is considered to be present if there is a peak in the PSD that is several times above the background noise with a confidence level over 95\%. The background noise is well fitted to a combination of red noise and white noise.
The proposed method has already been successfully applied to other observations, but never to H$\alpha$ data. 

Filament oscillations consist of periodic displacements of its cold plasma over the solar disc. These oscillations do not have to produce intrinsic oscillations of their emission. However, the periodic motion of the cold plasma produces periodic occultations of the emission from the chromosphere under and around the filament. This allows us to apply the spectral technique for our purpose. To understand how this technique is applied, we reproduce the oscillation detection process by mimicking the oscillations of filaments in H$\alpha$.
We analyze the PSD of each pixel in the domain and see that around the filament the PSD has a strong peak centred on the period of the oscillation. 

We applied the new technique to some events already reported in the \citet{luna_gong_2018} catalogue. In this way, we have been able to verify the reliability of the new technique.
The first case study corresponds to the first event in the catalogue. We detect a strong signal around the same filament that appears in the catalogue and with a similar period. Moreover, the regions where a PSD above the detection threshold is detected are on the path used in the catalogue to track the oscillations. The intensity in the regions with positive detection shows clear periodic fluctuations. 
The second case we consider is a faint filament near the limb where the chromospheric intensity decreases as we approach the edge of the disc (limb-darkening effect). We found that the method detects oscillations in the same region of the filament as in the catalogue and with a similar period.
The third case consists of a very clear oscillation with many oscillation cycles.
In this case, we see that the visual oscillation tracking method and the automatic oscillation tracking method disagree. The regions with positive oscillation are not aligned with the slit used in the catalogue. This suggests that the new method is more accurate in determining the regions with periodic motions.
We then show two opposite cases: one with a very small amplitude and one with a very large amplitude. The first one belongs to the catalogue but the second one has not been described in the literature but its oscillations are very clear. 
In the ROI of the first case, we detected the oscillation described in the catalogue but we also found periodic movements in a second filament in the same area. However, the oscillation period does not coincide with both techniques. Furthermore, the regions where we detected the oscillation are not on the slit used in the catalogue. This could indicate that the slit used was not optimal for tracking the oscillation explaining the discrepancy with the automatic method.
The opposite case corresponds to a large oscillation after a partial eruption and a flare. The method detects this oscillation perfectly.
Finally, we consider an arbitrary day and analyze the whole solar disc where we do not know if there are a priori oscillations. We detect five zones with oscillations between 51 and 69 minutes. All oscillations correspond to periodic filament motions except one which corresponds to a quasi-periodically recurring surge-like activity. Future research will investigate a method that allows us to distinguish between quasi-periodic fluctuations in intensity associated with e.g. repetitive jet activity from actual oscillations in filaments.

In this first study, we focused on frequencies lower than 1 mHz. The reason is that for frequencies above 1 mHz many small scattered areas pass the detection criterion. These may be due to false positives or actual chromospheric fluctuations. In future research, we will investigate filament oscillations at higher frequencies in more detail.

This work consists of a proof of concept for automatically detecting oscillations in solar filaments. We prove that the presented spectral method is very powerful for detecting these periodic motions and that it can be used to analyze solar data on a massive scale. 
However, for this end modifications to the method as presented in \S\ref{sec:data-processing} need to be made.
The region of interest should consist of the entire Sun and not just a portion of the solar disk. 
In addition, we should combine the data from all telescopes of the GONG network to have long temporal sequences. This will allow us to have a larger spectral resolution and also to be able to study very long period oscillations such as the very long periods of several tens of hours \citep{foullon_detection_2004,efremov_ultra_2016}. 
This study has shown that it is possible to detect oscillation in filaments and their periods. In future research, we will study the possibility of parameterising the damping time, the velocity amplitude and the direction of oscillation. 
These are only part of the challenges that will have to be addressed in order to develop in the future an automated code for the detection and parameterisation of oscillations massively in solar filaments.
With such a code we could understand how the oscillations in solar filaments evolve over several solar cycles using data from GONG or any other H$\alpha$ data repository. This research is also relevant for the planned next-generation GONG \citep[ngGONG,][]{hill_nggong_2019} network of telescopes.

\begin{acknowledgements}
M.L. acknowledges support through the Ram\'on y Cajal
fellowship RYC2018-026129-I from the Spanish Ministry of
Science and Innovation, the Spanish National Research Agency
(Agencia Estatal de Investigaci\'on), the European Social Fund
through Operational Program FSE 2014 of Employment,
Education and Training and the Universitat de les Illes Balears. This publication is part of the R+D+i project PID2020-112791GB-I00, financed by MCIN/AEI/10.13039/501100011033. The authors also acknowledges support from the International Space Sciences Institute (ISSI) via team 413 on ``Large-Amplitude Oscillations as a Probe of Quiescent and Erupting Solar Prominences.''
\end{acknowledgements}


\begin{thebibliography}{38}
\expandafter\ifx\csname natexlab\endcsname\relax\def\natexlab#1{#1}\fi

\bibitem[{Arregui {et~al.}(2018)Arregui, Oliver, \&
  Ballester}]{arregui_prominence_2018}
Arregui, I., Oliver, R., \& Ballester, J.~L. 2018, Living Reviews in Solar
  Physics, 15, 3, publisher: Springer International Publishing

\bibitem[{Auch{\`e}re {et~al.}(2014)Auch{\`e}re, Bocchialini, Solomon, \&
  Tison}]{auchere_long-period_2014}
Auch{\`e}re, F., Bocchialini, K., Solomon, J., \& Tison, E. 2014, Astronomy and
  Astrophysics, 563, A8, publisher: EDP Sciences

\bibitem[{Auch{\`e}re {et~al.}(2016)Auch{\`e}re, Froment, Bocchialini, Buchlin,
  \& Solomon}]{auchere_fourier_2016}
Auch{\`e}re, F., Froment, C., Bocchialini, K., Buchlin, E., \& Solomon, J.
  2016, The Astrophysical Journal, 825, 110, publisher: IOP Publishing

\bibitem[{Bashkirtsev \& Mashnich(1993)}]{bashkirtsev_some_1993}
Bashkirtsev, V.~S. \& Mashnich, G.~P. 1993, Astronomy and Astrophysics, 279,
  610, publisher: EDP Sciences

\bibitem[{Bernasconi {et~al.}(2005)Bernasconi, Rust, \&
  Hakim}]{bernasconi_advanced_2005}
Bernasconi, P.~N., Rust, D.~M., \& Hakim, D. 2005, Solar Physics, 228, 97,
  publisher: Springer Netherlands

\bibitem[{Bi {et~al.}(2014)Bi, Jiang, Yang, Hong, Li, Yang, \&
  Yang}]{bi_solar_2014}
Bi, Y., Jiang, Y., Yang, J., {et~al.} 2014, The Astrophysical Journal, 790, 100

\bibitem[{Bonnin {et~al.}(2013)Bonnin, Aboudarham, Fuller, Csillaghy, \&
  Bentley}]{bonnin_automation_2013}
Bonnin, X., Aboudarham, J., Fuller, N., Csillaghy, A., \& Bentley, R. 2013,
  Solar Physics, 283, 49, publisher: Springer Netherlands

\bibitem[{Chandra {et~al.}(2017)Chandra, Filippov, Joshi, \&
  Schmieder}]{chandra_two-step_2017}
Chandra, R., Filippov, B., Joshi, R., \& Schmieder, B. 2017, Solar Physics,
  292, 81

\bibitem[{Coffey \& Hanchett(1998)}]{coffey_digital_1998}
Coffey, H.~E. \& Hanchett, C.~D. 1998, 150, 488, conference Name: IAU Colloq.
  167: New Perspectives on Solar Prominences ADS Bibcode: 1998ASPC..150..488C

\bibitem[{Efremov {et~al.}(2016)Efremov, Parfinenko, \&
  Solov'ev}]{efremov_ultra_2016}
Efremov, V.~I., Parfinenko, L.~D., \& Solov'ev, A.~A. 2016, Solar Physics, 291,
  3357, publisher: Springer Netherlands

\bibitem[{Foullon {et~al.}(2004)Foullon, Verwichte, \&
  Nakariakov}]{foullon_detection_2004}
Foullon, C., Verwichte, E., \& Nakariakov, V.~M. 2004, Astronomy and
  Astrophysics, 427, L5, publisher: EDP Sciences

\bibitem[{Fuller {et~al.}(2005)Fuller, Aboudarham, \&
  Bentley}]{fuller_filament_2005}
Fuller, N., Aboudarham, J., \& Bentley, R.~D. 2005, Solar Physics, 227, 61,
  publisher: Springer Netherlands

\bibitem[{Gao {et~al.}(2002)Gao, Wang, \& Zhou}]{gao_development_2002}
Gao, J., Wang, H., \& Zhou, M. 2002, Solar Physics, 205, 93, publisher:
  Springer Netherlands

\bibitem[{Gregory(2001)}]{gregory_bayesian_2001}
Gregory, P.~C. 2001, 568, 557, conference Name: Bayesian Inference and Maximum
  Entropy Methods in Science and Engineering

\bibitem[{Hao {et~al.}(2015)Hao, Fang, Cao, \& Chen}]{hao_statistical_2015}
Hao, Q., Fang, C., Cao, W., \& Chen, P.~F. 2015, The Astrophysical Journal
  Supplement Series, 221, 33, publisher: IOP Publishing

\bibitem[{Hershaw {et~al.}(2011)Hershaw, Foullon, Nakariakov, \&
  Verwichte}]{hershaw_damped_2011}
Hershaw, J., Foullon, C., Nakariakov, V.~M., \& Verwichte, E. 2011, Astron.
  Astrophys., 531, A53

\bibitem[{Hill {et~al.}(2019)Hill, Hammel, Martinez-Pillet, Wijn, Gosain,
  Burkepile, Henney, McAteer, Bain, Manchester, Lin, Roth, Ichimoto, \&
  Suematsu}]{hill_nggong_2019}
Hill, F., Hammel, H., Martinez-Pillet, V., {et~al.} 2019, Bulletin of the AAS,
  51

\bibitem[{Hyder(1966)}]{hyder_winking_1966}
Hyder, C. 1966, Z. Astrophys., 63, 78

\bibitem[{Ireland {et~al.}(2010)Ireland, Marsh, Kucera, \&
  Young}]{ireland_automated_2010}
Ireland, J., Marsh, M.~S., Kucera, T.~A., \& Young, C.~A. 2010, Solar Physics,
  264, 403, publisher: Springer Netherlands

\bibitem[{Jing {et~al.}(2004)Jing, Yurchyshyn, Yang, Xu, \&
  Wang}]{jing_relation_2004}
Jing, J., Yurchyshyn, V.~B., Yang, G., Xu, Y., \& Wang, H. 2004, The
  Astrophysical Journal, 614, 1054, publisher: IOP Publishing

\bibitem[{Kleczek \& Kuperus(1969)}]{kleczekkuperus_oscillatory_1969}
Kleczek, J. \& Kuperus, M. 1969, Solar Phys., 6, 72

\bibitem[{Leibacher {et~al.}(2010)Leibacher, Sakurai, Schrijver, \& van
  Driel-Gesztelyi}]{leibacher_solar_2010}
Leibacher, J., Sakurai, T., Schrijver, C.~J., \& van Driel-Gesztelyi, L. 2010,
  Solar Physics, 263, 1, aDS Bibcode: 2010SoPh..263....1L

\bibitem[{Luna {et~al.}(2018)Luna, Karpen, Ballester, Muglach, Terradas,
  Kucera, \& Gilbert}]{luna_gong_2018}
Luna, M., Karpen, J., Ballester, J.~L., {et~al.} 2018, The Astrophysical
  Journal Supplement Series, 236, 35, publisher: IOP Publishing

\bibitem[{Luna {et~al.}(2014)Luna, Knizhnik, Muglach, Karpen, Gilbert, Kucera,
  \& Uritsky}]{luna_observations_2014}
Luna, M., Knizhnik, K., Muglach, K., {et~al.} 2014, The Astrophysical Journal,
  785, 79, publisher: IOP Publishing

\bibitem[{Luna {et~al.}(2017)Luna, Su, Schmieder, Chandra, \&
  Kucera}]{luna_large-amplitude_2017}
Luna, M., Su, Y., Schmieder, B., Chandra, R., \& Kucera, T.~A. 2017, The
  Astrophysical Journal, 850, 143, publisher: IOP Publishing

\bibitem[{Mackay(2014)}]{mackay_formation_2014}
Mackay, D.~H. 2014, in Solar {Prominences}, 415th edn. (Cham: Solar
  Prominences), 355--380

\bibitem[{McIntosh(1972)}]{mcintosh_solar_1972}
McIntosh, P.~S. 1972, Reviews of Geophysics, 10, 837, \_eprint:
  https://onlinelibrary.wiley.com/doi/pdf/10.1029/RG010i003p00837

\bibitem[{McIntosh {et~al.}(2008)McIntosh, De~Pontieu, \&
  Tomczyk}]{mcintosh_coherence-based_2008}
McIntosh, S.~W., De~Pontieu, B., \& Tomczyk, S. 2008, Solar Physics, 252, 321

\bibitem[{Minarovjech {et~al.}(1998)Minarovjech, Rybansk{\'y}, \& Ru{\v
  s}in}]{minarovjech_prominences_1998}
Minarovjech, M., Rybansk{\'y}, M., \& Ru{\v s}in, V. 1998, Solar Physics, 177,
  357

\bibitem[{Nakariakov \& King(2007)}]{nakariakov_coronal_2007}
Nakariakov, V.~M. \& King, D.~B. 2007, Solar Physics, 241, 397

\bibitem[{Shih \& Kowalski(2003)}]{shih_automatic_2003}
Shih, F.~Y. \& Kowalski, A.~J. 2003, Solar Physics, 218, 99, publisher:
  Springer Netherlands

\bibitem[{Terradas {et~al.}(2002)Terradas, Molowny-Horas, Wiehr, Balthasar,
  Oliver, \& Ballester}]{terradas_two-dimensional_2002}
Terradas, J., Molowny-Horas, R., Wiehr, E., {et~al.} 2002, Astronomy and
  Astrophysics, 393, 637, publisher: EDP Sciences

\bibitem[{Torrence \& Compo(1998)}]{torrence_practical_1998}
Torrence, C. \& Compo, G.~P. 1998, Bulletin of the American Meteorological
  Society, 79, 61

\bibitem[{{van Ballegooijen}(2004)}]{van-ballegooijen_observations_2004}
{van Ballegooijen}, A.~A. 2004, \apj, 612, 519

\bibitem[{VanderPlas(2018)}]{vanderplas_understanding_2018}
VanderPlas, J.~T. 2018, The Astrophysical Journal Supplement Series, 236, 16,
  publisher: American Astronomical Society

\bibitem[{Zechmeister \& K{\"u}rster(2009)}]{zechmeister_generalised_2009}
Zechmeister, M. \& K{\"u}rster, M. 2009, Astronomy \& Astrophysics, 496, 577

\bibitem[{Zhang {et~al.}(2012)Zhang, Chen, Xia, \&
  Keppens}]{zhang_observations_2012}
Zhang, Q.~M., Chen, P.~F., Xia, C., \& Keppens, R. 2012, Astronomy \&
  Astrophysics, 542, A52, publisher: EDP Sciences

\bibitem[{{Zharkova} {et~al.}(2004){Zharkova}, {Aboudarham}, {Zharkov},
  {Ipson}, {Benkhalil}, \& {Fuller}}]{zharkova_seachable_2004}
{Zharkova}, V.~V., {Aboudarham}, J., {Zharkov}, S.~I., {et~al.} 2004, in AGU
  Fall Meeting Abstracts, Vol. 2004, SH52A--04

\end{thebibliography}

\end{document}